\documentclass[sigconf]{acmart}

\usepackage{xcolor}

\AtBeginDocument{%
  }

\copyrightyear{2026}
\acmYear{2026}
\setcopyright{cc}
\setcctype{by}
\acmConference[CHI '26]{Proceedings of the 2026 CHI Conference on Human Factors in Computing Systems}{April 13--17, 2026}{Barcelona, Spain}
\acmBooktitle{Proceedings of the 2026 CHI Conference on Human Factors in Computing Systems (CHI '26), April 13--17, 2026, Barcelona, Spain}
\acmPrice{}
\acmDOI{10.1145/3772318.3791314}
\acmISBN{979-8-4007-2278-3/2026/04}

\newcommand{\sys}{CHOIR}

\newcommand{\rnr}[1]{#1}

\newcommand{\newadd}[1]{#1} 

\definecolor{feature1}{HTML}{3DADFF} 
\definecolor{feature2}{HTML}{F24822} 
\definecolor{feature3}{HTML}{FF9E42} 
\definecolor{feature4}{HTML}{5BBE69} 

\newcommand{\bluecirclewrap}[1]{{\sffamily\textcolor{feature1}{\Large\textcircled{\normalsize\textbf{#1}}}}}
\newcommand{\redcirclewrap}[1]{{\sffamily\textcolor{feature2}{\Large\textcircled{\normalsize\textbf{#1}}}}}
\newcommand{\orangecirclewrap}[1]{{\sffamily\textcolor{feature3}{\Large\textcircled{\normalsize\textbf{#1}}}}}
\newcommand{\greencirclewrap}[1]{{\sffamily\textcolor{feature4}{\Large\textcircled{\normalsize\textbf{#1}}}}}

\usepackage{enumitem}
\usepackage{stfloats}

\definecolor{selectionboxcolor}{HTML}{f0f0f0}

\begin{document}

\title{\system: A Chatbot-mediated Organizational Memory Leveraging Communication in University Research Labs}

\author{Sangwook Lee}
\orcid{0000-0002-2600-4769}
\affiliation{%
  \institution{Virginia Tech}
  \city{Blacksburg}
  \state{Virginia}
  \country{USA}
}
\email{sangwooklee@vt.edu}

\author{Adnan Abbas}
\orcid{0009-0005-8728-875X}
\affiliation{%
  \institution{Virginia Tech}
  \city{Blacksburg}
  \state{Virginia}
  \country{USA}
}
\email{adnana99@vt.edu}

\author{Yan Chen}
\orcid{0000-0002-1646-6935}
\affiliation{%
  \institution{Virginia Tech}
  \city{Blacksburg}
  \state{Virginia}
  \country{USA}
}
\email{ych@vt.edu}

\author{Young-Ho Kim}
\orcid{0000-0002-2681-2774}
\affiliation{%
  \institution{NAVER AI Lab}
  \city{Seongnam}
  \country{Republic of Korea}
}
\email{yghokim@younghokim.net}

\author{Sang Won Lee}
\orcid{0000-0002-1026-315X}
\authornote{Sang Won Lee conducted this work as a visiting scholar at NAVER AI Lab.}
\affiliation{%
  \institution{Virginia Tech}
  \city{Blacksburg}
  \state{Virginia}
  \country{USA}
}
\email{sangwonlee@vt.edu}

\newcommand{\system}{CHOIR}

\begin{abstract}
University research labs often rely on chat-based platforms for communication and project management, where valuable knowledge surfaces but is easily lost in message streams. Documentation can preserve knowledge, but it requires ongoing maintenance and is challenging to navigate. Drawing on formative interviews that revealed organizational memory challenges in labs, we designed CHOIR, an LLM-based chatbot that supports organizational memory through four key functions: document-grounded Q\&A, Q\&A sharing for follow-up discussion, knowledge extraction from conversations, and AI-assisted document updates. We deployed CHOIR in four research labs for one month (n=21), where the lab members asked 107 questions and lab directors updated documents 38 times in the organizational memory. Our findings reveal a privacy-awareness tension: questions were asked privately, limiting directors' visibility into documentation gaps. Students often avoided contribution due to challenges in generalizing personal experiences into universal documentation. We contribute design implications for privacy-preserving awareness and supporting context-specific knowledge documentation.
\end{abstract}



\begin{teaserfigure}
  \centering
  \includegraphics[width=\textwidth]{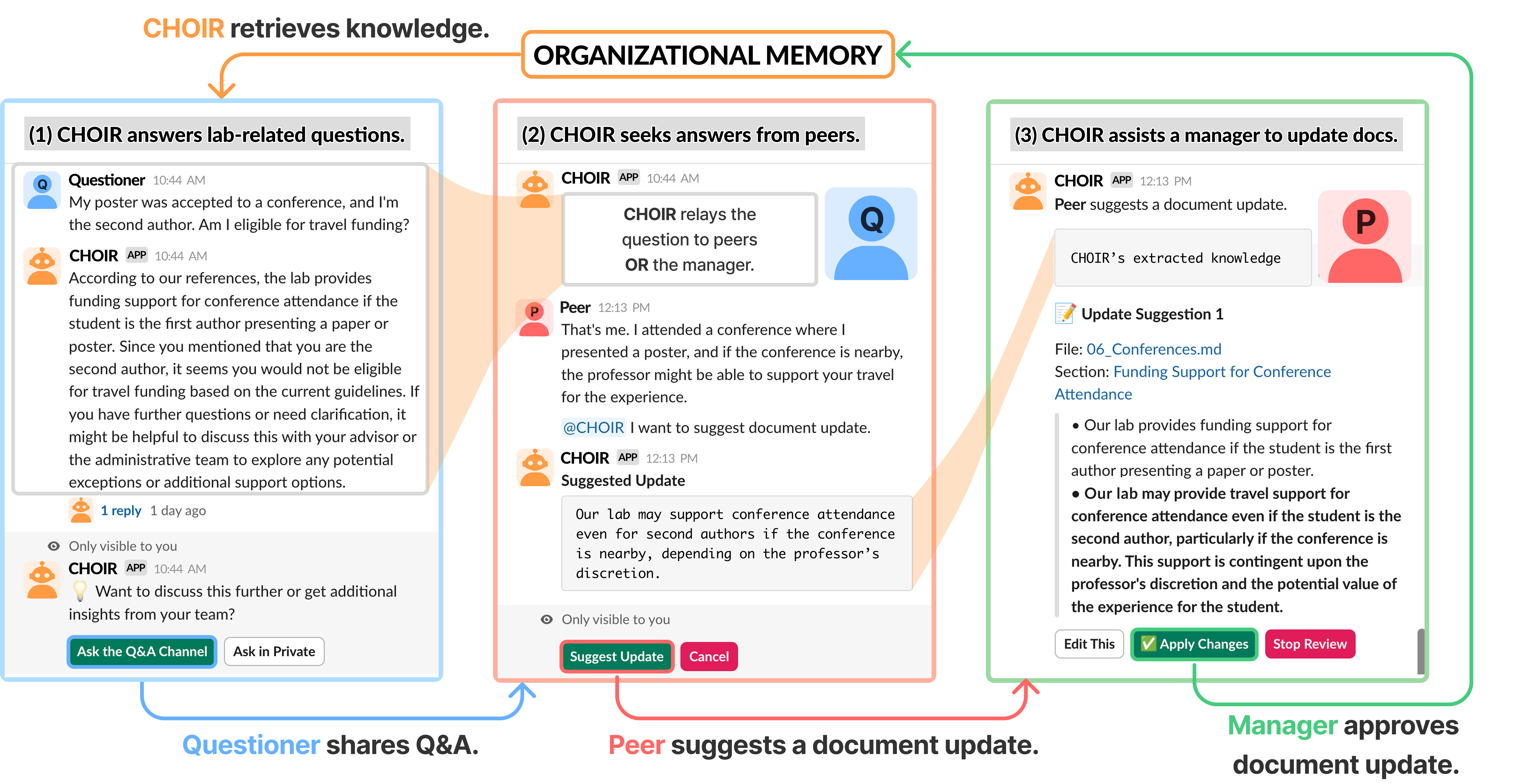}
  \caption{CHOIR's workflow across different communication channels}
  \Description{This figure illustrates CHOIR's workflow through three sequential interactions across different communication channels. Panel (1) shows a Questioner in a direct message with CHOIR asking about conference travel funding eligibility and receiving a document-grounded answer with source references, with options to share in the Q&A channel or ask privately. Panel (2) depicts a Peer in the public Q&A channel providing additional conference experience and suggesting a document update after CHOIR relays the question to peers or the manager. Panel (3) shows a Manager in a direct message with CHOIR receiving extracted knowledge and document update suggestions about conference funding support, with options to edit and apply changes. The bottom labels indicate how each role contributes to organizational memory: Questioner initiates inquiries, Peer provides collaborative input, and Manager approves document updates.}
  \label{fig:system_teaser}
\end{teaserfigure}

\maketitle
\section{Introduction}

Modern university research labs increasingly rely on workplace instant-messaging platforms such as Slack and Microsoft Teams to facilitate communication and collaboration among lab members, including those working remotely~\cite{perkel2017how, wang2022group}. Slack alone reported over 1.2 million users across more than 3,000 higher education institutions during the COVID-19 pandemic in 2020~\cite{slack2020distance}. These platforms serve as channels through which lab members share knowledge crucial and practical to their work—e.g., academic writing tips, internship opportunities, and travel grant information—on an ongoing basis~\cite{isaacs2002character,anders2016team,wang2022group}.
However, this knowledge is often lost. 
Messages are stored chronologically and without semantic structure, making it difficult to retrieve relevant information once it is buried in long conversation threads~\cite{zhang2017wikum,zhang2018making, pasad2020understanding, 10.1145/3132390.3132392}. 
As a result, lab directors may receive the same questions repeatedly or give the same advice repeatedly, e.g., correcting punctuation in latex. Students are left to learn through trial and error~\cite{anderson1996situated}. 

Prior literature has suggested addressing this challenge through sociotechnical interventions that establish an \textit{organizational memory}, which documents and preserves knowledge accumulated by an organization over time~\cite{walsh1991organizational, ackerman2004organizational}. Some research labs implement such organizational knowledge systems using collaborative documents such as Google Docs~\cite{sun2014collaboration} or group wikis~\cite{wagner2004wiki,grudin2010wikis}, often in the form of lab handbooks or policy documents. However, maintaining these resources requires substantial manual effort~\cite{orlikowski1992learning,lethbridge2003how, holtzblatt2010factors, koivisto2025pitfalls}. In practice, only professors find it difficult to consistently maintain such documents~\cite{star1999layers,sun2014collaboration}—a finding also confirmed by our formative study. As a result, much of the knowledge remains undocumented, residing instead in the memory of professors and lab members~\cite{droege2003employee,galan2023knowledge}. Even when central documents exist, students often struggle to read them entirely or locate the right information when they need it~\cite{boardman2004stuff, whittaker2011personal}. Consequently, they must turn to peers or professors, making knowledge transfer inefficient.

To address this challenge, we present \sys{} (Chat-based Helper for Organizational Intelligence Repository), a chatbot-based organizational memory management system embedded in Slack. 
The primary goal of \sys{} is twofold: (1) to build and sustain a research lab's organizational memory, and (2) to enable lab members to easily access knowledge from it. To that end, our approach embeds organizational memory directly within their communication platform, such as Slack, allowing students to ask questions to a chatbot and allowing professors to modify documents within the platform. 
Specifically, \sys{} integrates four interconnected interactions: \textbf{Document-grounded Q\&A}, which retrieves knowledge with relevant references from a collection of documents when a user asks a question; \textbf{Q\&A Sharing for Follow-up Discussion}, which allows users to share Q\&A with peers to address unanswered questions or extend discussions; \textbf{Knowledge Extraction from Conversation}, which captures new knowledge that naturally emerges from Slack messages; and \textbf{AI-assisted Document Update}, which facilitates document maintenance by helping managers locate relevant documents and suggest edits. In this way, \sys{} supports an organizational memory practice that leverages lab members’ communication in Slack to retrieve and sustain knowledge, reducing lab directors' burden of continuously updating documents while providing members with immediate and convenient access to organizational memory.
While this approach can be potentially applicable to many different types of organizations, such as an online community and a company, we chose research labs as our target environment because they constantly experience member turnover, rely on regular communication, and maintain various types of documentation that require collaborative updates~\cite{tuma2021dark}.

To examine both current challenges of organizational memory for university research labs and the opportunities of a chatbot-driven organizational memory, we aim to answer the following two questions:
\vspace{-5pt}
\begin{itemize}
    \item \textbf{RQ1}: What are the current challenges of managing and utilizing organizational memory in research labs?
    \item \textbf{RQ2}: How do lab members engage with and experience a chatbot-driven organizational memory for managing and utilizing organizational memory in their online communication?
\end{itemize}
\vspace{-5pt}

We followed a three-stage approach: (1) a formative study with faculty and graduate students (N=15) to understand their organizational memory practice, if any, challenges in transferring and retrieving knowledge among lab members; (2) the design and implementation of \sys{} guided by these findings; and (3) a one-month field deployment (N=21) across four university research labs, preceded by document collection and structured onboarding.
Based on our interaction trace analysis and follow-up interviews after the field study, we found that \sys{} helped students obtain answers available in the documents. In addition, it supported professors with no prior experience in maintaining organizational memory to document new knowledge that students wanted to learn or needed to know when questions were shared publicly. \rnr{However, our findings revealed a fundamental tension between students' privacy preferences and directors' need for awareness of knowledge gaps.} Students faced psychological barriers to sharing their questions—either from concerns about bothering their peers and professors, or from reluctance due to how others might perceive them for not knowing the answers to seemingly basic questions. Similar barriers were noted for answering other people's questions \rnr{and for contributing to documentation updates}.
Based on the findings, we discuss the design implications for AI-mediated organizational memory systems, including \rnr{the privacy-awareness trade-off, the role of AI as a facilitator for human communication, and approaches to expand} the perceived boundary of documentable knowledge.

The research contributions of this paper include: 

\vspace{-5pt}
\begin{enumerate}
    \item  An empirical investigation of the challenges surrounding organizational memory practices in university research labs.
    \item  A socio-technical workflow that integrates conversational Q\&A, knowledge sharing, and AI-assisted documentation to support organizational memory.
    \item Findings from a month-long deployment that reveal real-world usage patterns, challenges and design implications for AI-mediated organizational memory systems.
\end{enumerate}

\section{Related Work}
\newadd{While prior work on personal knowledge management~\cite{dumais2003stuff, gemmell2006mylifebits, capra2005using} focused on search-based retrieval across collected sources, an organization's knowledge management is challenging due to its scale and group dynamics.}
We explore prior research on organizational memory and knowledge management systems, examining how digital platforms have evolved to support knowledge preservation and sharing. We then investigate approaches to transforming conversational data into organizational knowledge, and finally review the emerging role of chatbots and large language models (LLMs) in facilitating knowledge work.

\subsection{From Repository Models to Distributed Organizational Memory}
Organizational memory (OM), defined as stored information from an organization's history that can be brought to bear on present decisions~\cite{walsh1991organizational}, has been a central concern in knowledge management research. Early approaches assumed that organizational knowledge could be systematically captured, stored, and retrieved through repository-based systems. Initial groupware systems attempted to implement this vision by extracting knowledge from individuals and storing it in centralized databases. However, deployments revealed fundamental social and organizational barriers that overshadowed technical challenges. Research identified that knowledge sharing requires addressing incentive structures~\cite{orlikowski1992learning,star1999layers}, organizational power dynamics~\cite{grudin1994groupware}, and the scattered nature of informal knowledge that often remains undocumented~\cite{conklin1997designing}. Competitive organizational cultures frequently inhibited knowledge sharing, as employees treated expertise as personal assets.

Meanwhile, a different approach emerged with Ackerman's Answer Garden series, which explored hybrid human-system methods. Answer Garden~\cite{ackerman1990answer} allowed users to navigate branching questions and automatically routed unanswered queries to experts, who could then contribute responses to the knowledge base. Answer Garden 2~\cite{ackerman1996answer} introduced cascading routing from peers to specialists to experts. However, field studies revealed persistent challenges~\cite{ackerman1998augmenting}: mismatched expectations between experts and users regarding answer detail, user reluctance to burden experts, and, in industrial deployments, conflicts between organizational silos and departmental politics~\cite{pipek2003pruning}. These findings led to reconceptualizing OM as a distributed cognitive system rather than a centralized repository. Ackerman~\cite{ackerman1998considering} demonstrated through ethnographic research that OM encompasses individual memory, peer knowledge, documents, and customer interactions in interconnected forms. Knowledge crossing organizational boundaries transforms into boundary objects~\cite{star1989institutional}, requiring recontextualization for effective use~\cite{ackerman2004organizational}. 

With the advancement of web technologies, wiki systems demonstrated this potential through collaborative editing~\cite{wagner2004wiki}, but organizational deployments encountered familiar social barriers. Danis~\cite{danis2008wiki} deployed ResearchWiki in a research organization but found limited participation due to competitive culture and modification anxiety. \rnr{Similarly, Munson~\cite{10.1145/1822258.1822283} found in a MediaWiki deployment that while the system motivated documentation of past projects and enabled discovery of organizational work, members felt uncomfortable making substantial edits to others' content. Broader studies of enterprise wikis found that making workers participate in wiki is challenging and organization's culture that support collaboration is necessary~\cite{kiniti2013wikis}.} Subsequent research identified systemic barriers: additional effort requirements, information sensitivity concerns, discomfort with unfinished work publication, and openness anxiety~\cite{holtzblatt2010factors}. The collaborative editing model effective in public contexts created stress in organizational environments where authorship and accountability mattered~\cite{grudin2010wikis}. Alternative platforms like Google Docs achieved better adoption through simplified user experiences that naturally facilitated role differentiation without wiki complexity~\cite{sun2014collaboration}, with supporting technologies focusing on transparent collaboration history to navigate social complexities~\cite{wang2015docuviz}.

\subsection{Leveraging Conversational Platforms for Knowledge Management}
Researchers have increasingly recognized conversations as valuable sources of organizational knowledge, particularly for capturing tacit knowledge that resists formalization~\cite{polanyi1997tacit}. The challenge is transforming conversational knowledge into explicit, reusable forms while preserving contextual richness~\cite{nonaka1998knowledgecreating}. Early digital communication platforms revealed this potential, with email evolving beyond simple messaging to serve multiple organizational functions, including task management, document archiving, and conversation tracking, though leading to information overload~\cite{whittaker1996email}. Instant messaging further demonstrated how informal channels could become knowledge repositories for relationship maintenance and work coordination~\cite{nardi2000interaction,isaacs2002character,handel2002what}.

As diverse platforms began supporting interpersonal conversations, researchers discovered that various platforms were serving knowledge management roles. Platform diversification revealed knowledge management opportunities in tools not originally designed for that purpose. Software issue trackers evolved into OM and coordination systems~\cite{bertram2010communication}, while corporate social networking enabled help requests and work coordination through informal interactions~\cite{ehrlich2010microblogging}. Research showed these platforms improved employees' metaknowledge awareness through social mediation~\cite{leonardi2015ambient,mamykina2011design,nisar2019social}. Team communication platforms like Slack provided unprecedented conversation visibility and multi-threaded collaboration~\cite{anders2016team}, but also introduced information overload and conversational inefficiencies from simultaneous topic progression~\cite{pasad2020understanding}. In research environments, Slack served as both a knowledge-sharing hub and a culture formation space~\cite{perkel2017how}, with conversational data proving valuable for tasks like API recommendations from developer discussions~\cite{chatterjee2019exploratory}.

Despite these discoveries, converting conversations into persistent knowledge required addressing the challenge of preserving natural communication flow while enabling knowledge extraction. Early approaches focused on visualization and threading for improved retrieval~\cite{donath1999visualizing,kerr2003thread,faridani2010opinion}, but required manual navigation. More sophisticated systems automated knowledge extraction through annotation~\cite{ackerman2003idiag,fono2006structuring} and collaborative summarization~\cite{zhang2017wikum}. Chat-based knowledge management systems achieved greater sophistication with Tilda, which automatically summarizes tagged conversations~\cite{zhang2018making}, and tools combining topic learning with personalized summaries~\cite{tepper2018collabot}. \newadd{Social Q\&A systems demonstrated how questions can surface knowledge gaps and how answers gradually accumulate into shared resources~\cite{richardson2011supporting}.} QWiki demonstrated that Q\&A and documentation approaches could be mutually reinforcing in MOOC environments~\cite{setia2020qwiki}. This trajectory confirms Ackerman's~\cite{ackerman2013sharing} prediction that OM should move toward communication-centric, people-oriented approaches, while raising questions about scalability and sustainability.

\subsection{AI-Powered Knowledge Management: From Chatbots to LLMs}
These scalability and sustainability challenges led to the emergence of AI-powered approaches to address the limitations of manual conversational knowledge extraction. As demonstrated by early systems like Collabot~\cite{tepper2018collabot} and Tilda~\cite{zhang2018making}, many of these solutions took the form of task-based chatbots~\cite{adamopoulou2020overview}. These systems provided basic topic extraction and unidirectional summarization capabilities. Chatbots evolved from passive observers to active conversation participants, taking specialized roles as archivists, social organizers, and moderators~\cite{seering2019dyadic,kim2020bot,kim2021moderator}. Research showed users felt more comfortable sharing knowledge with chatbots than humans, reducing social barriers~\cite{lee2020hear}.

With the advancement of chatbot technologies, LLMs enabled sophisticated conversational knowledge processing beyond basic pattern recognition. LLMs demonstrated superior performance in email and meeting summarization~\cite{zhang2021emailsum,zhong2021qmsum} and enabled interactive refinement where users could update AI-generated summaries for improved performance~\cite{asthana2025summaries}. Retrieval-Augmented Generation (RAG) addressed the challenge of grounding conversational AI in organizational knowledge bases~\cite{lewis2021retrievalaugmented}, with RAG systems searching existing documents and incorporating relevant information into responses, improving accuracy and reducing hallucination~\cite{gao2024retrievalaugmented}. Beyond retrieval and summarization, LLMs enable dynamic document updating based on conversational knowledge~\cite{laban2024chat} and facilitate complex group interactions~\cite{chiang2024enhancing}.

\subsection{Summary and Connection to CHOIR}
This review reveals consistent evolution in OM systems: from repository models to distributed cognitive systems, from formal documentation to conversational capture, and from manual curation to AI-assisted transformation. However, persistent challenges have remained consistent across decades—incentive misalignment, expert burden, participation barriers, and information overload—indicating that technical advances alone are insufficient. \rnr{Prior AI-based systems such as Tilda and Collabot primarily address subprocesses of the OM lifecycle, mainly capturing and summarizing conversation~\cite{zhang2018making, tepper2018collabot}, but do not support the full cycle: how extracted knowledge is validated, integrated into OM documents, and easily retrieved by the members.} \sys{} addresses this gap by synthesizing three research streams: embedding OM in existing social systems, leveraging existing conversational interactions, and exploiting LLMs' capabilities. By operating within Slack where research teams already communicate, CHOIR avoids traditional adoption barriers while implementing document-grounded Q\&A, conversational knowledge extraction, and AI-assisted documentation workflows. The system design incorporates lessons from prior research: mutually reinforcing Q\&A and documentation cycles, integration with existing communication flows, and AI assistance that reduces rather than increases user burden. CHOIR thus represents both technological synthesis and an approach to addressing the persistent social challenges that have limited OM system adoption in collaborative research environments.

\section{Formative Study}
To address RQ1, i.e., understanding challenges of organizational memory in research labs, we conducted a formative study with university research lab members, both professors and students.

\subsection{Method}

We recruited participants through a screening survey distributed to faculty and graduate students at research-oriented universities. We advertised the study on the author's social media to recruit professors in other research-oriented universities.
We recruited 15 participants: 11 faculty members (P1-P11) from various academic disciplines, primarily Computer Science (7) but also Industrial Engineering (2), Information Science (1), and Human Development and Family Science (1), and 4 graduate students (S1-S4) from different disciplines. We targeted professors who directed research labs with more than 4 Ph.D. students and had experience or interest in maintaining organizational memory. 
Faculty participants had been running their research labs for an average of 6.4 years (SD = 3.78, range: 4-13 years) and supervised an average of 5 Ph.D. students (SD = 1.67, range: 2-7 students).
Some of the participants had a well-established organizational memory practice, whereas others relied on more informal organizational methods.
This study was approved by the Virginia Tech Institutional Review Board (IRB 25-247).

We conducted semi-structured interviews lasting 45-60 minutes with each participant via Zoom. 
All interviews were 
audio-recorded with the participants' consent. The interview protocol explored the challenges that they face in addressing knowledge transfer and maintaining organizational memory, if any. 
We also presented early mock-up of \sys{} to elicit participants' reactions to possible features, rather than a fully implemented system. 
Participants were compensated with \$50 for faculty members and \$20 for graduate students upon completion.

Each interview was audio-recorded, and the first two authors transcribed the interview recordings, manually correcting the output of a transcription engine (Sonix AI\footnote{https://sonix.ai/}). After familiarization with the data, the second author performed a thematic analysis~\cite{clarke2014thematic}, and initially generated 214 codes. The three authors conducted axial coding by revising, merging, and deleting codes and identifying emergent themes. 

\subsection{Result}

Overall, both professors and students quickly recognized the benefits of documenting an organizational memory to help students access information. According to the professors interviewed, they often have to repeatedly address the same set of questions. These questions not only include administrative procedures like travel reimbursements or the Ph.D. qualification process but also involve insights or organizational memory gained over time, such as writing tips, information about equipment, research internships, and student expectations.

\begin{quote}
 \textit{``We have all kinds of gadgets, devices, and equipment. We have many robots, driving simulators, laptops, and eye trackers. And so they ask the same questions again and again.''} (P6)
\end{quote}

\begin{quote}
\textit{``Even if the questions are slightly different, I think the answers could be the same. [...] And so I sort of find myself having to repeat a lot of these things''} (P11)
\end{quote}

\noindent When professors lack formal documentation, they often rely on senior students to address recurring questions and facilitate knowledge transfer.  
However, they reported that, as senior students typically graduate after completing their degree, ensuring continuity in knowledge transfer becomes a separate challenge if not documented.

Of the 11 professors we interviewed, 7 maintained or had a document or a collection of documents that allowed students to find answers independently.
The level of detail in these documents also varied considerably, and most of them mentioned that they were not comprehensive or up-to-date. Some labs provided a two-page onboarding syllabus (P7), while others offered extensive and comprehensive documentation spanning over 80 pages (P8)

They also recognized that all the advice or tips they give in person is ``ephemeral'' and emphasized the benefits of documentation due to its persistent nature, allowing reuse and future reference.
However, our participants reported that utilizing and maintaining such documents was both challenging and concerning, which we will outline below. 

\subsubsection{\textbf{Challenges in Creating and Maintaining Lab Documentation}}

The primary challenge in documenting lab knowledge was the time-intensive nature of the task. 
Many participants admitted they were too ``busy” or ``lazy” to create such documents. 
In practice, documentation often takes a lower priority than more pressing responsibilities, such as research and teaching.

\begin{quote}
    \textit{``I think it's nice to have. I mean, I'm just lazy, and I don't have the time and energy, and it's just like, I'm not a type of person who could do that, but I think it's a nice resource to have. ''} (P3)
\end{quote}

\begin{quote}
\textit{``Documentation is usually one of those things that gets less attention, that gets deprioritized. [...] I have, I think, made 1 or 2 attempts over the years, but they never got fleshed out to the point where they would really be useful and worth sharing.''} (P9)
\end{quote}

\noindent While participants were concerned about the sheer amount of content they had to generate, keeping the document up to date with dynamic information was another significant challenge. As the document grows, identifying what needs to be updated and locating outdated information becomes increasingly difficult.

\begin{quote}
    \textit{``I go back to it [the lab handbook] and I look at it, I'm like, `Oh, wow. That's like no longer true.' [...] The more content I add, the more I have to maintain.''} (P8)
\end{quote}

\begin{quote}
    \textit{``There was one time there was like a whole step just completely missing, and I couldn't get it to work. [...] The people who had written the entry into the Notion had already graduated.''} (S4)
\end{quote}

\noindent These challenges are consistent with prior work on knowledge management; studies have shown the high cost of creating and maintaining knowledge artifacts~\cite{orlikowski1992learning, ackerman2013sharing}. Researchers have noted that documents tend to become outdated quickly, reducing their usefulness over time~\cite{grudin1994groupware,lethbridge2003how}. 

Based on these findings, we aim to design a system that facilitates the documentation of organizational memory. Our approach leverages frequently used communication platforms, such as Slack or Microsoft Teams, where lab directors or senior students answer questions and share tips. We can effectively capture and preserve various types of knowledge for documentation. The following design goals are derived from these findings.

\vspace{0.5em}
\noindent\textbf{Design Goal 1 (DG1):} Facilitate the process of documenting organizational memory, leveraging the existing conversation among lab members.

\subsubsection{\textbf{Need for Effective Knowledge Retreival}}

Another significant challenge raised by participants was the retrieval of knowledge. 
Even professors who maintained centralized documentation often struggled to ensure that students knew what information was available, where they are, and how to access it. 

\begin{quote}
    \textit{``Sometimes they will email me and ask me where the link is for the survey to report their hours [...] which for me is a little frustrating because it's linked in the syllabus for the lab.''} (P7)
\end{quote}

\begin{quote}
    \textit{``There is [sic] a lot of safety protocols that we need to go through. So if they don't pay too much attention during those [onboarding] meetings, then usually they struggle later on to find those resources or do things properly. ''} (S3)
\end{quote}

For the undocumented knowledge, similar challenges persist, if not worsen, as they often need to locate files, links, emails, or Slack messages to share with students. 
While they actively used search features, these tools were limited as information was scattered across multiple locations.

\begin{quote}
\textit{ ``Sometimes I'm frustrated that I have to go dig [the information] out and didn't have it nicely organized in one spot.''} (P5) 
\end{quote}

Communication platforms' features, such as search function or pinned messages, provide a way to reuse existing messages to some extent. However, they reported difficulties using those features for knowledge retrieval and management.

\begin{quote}
\textit{
``I probably don't use those [search functions] as effectively as I could. So, there's definitely been times when I'm trying to find an old conversation and I'm searching within a DM from maybe one student and it's not coming up, and I'm not really sure how I can find it.''} (P9)
\end{quote}

\begin{quote}
    \textit{``Pinning a message[...] it can get quite a large, like, a long list of items. And like the pinned message, you can only see, like, a couple.''} (P1)
\end{quote}

\noindent While existing functions allow users to retrieve or curate past messages they authored, this approach does not scale as the communication history grows, and retrieving information through pinned messages is similarly unsustainable.

Building on these findings, we propose designing a conversational agent that allows students to ask questions directly. The chatbot will respond using information available in relevant documents. This chat-based assistance will delegate the \textit{information search}, greatly reducing the effort required to retrieve information.
Additionally, we design the system to guide students to the original document and the specific section from which the answer was sourced, encouraging them to read the source directly. Providing a reference to the original document will help them understand the context of the answer and become familiar with the document. We summarize this approach in the following design goal.

\vspace{0.5em}
\noindent\textbf{Design Goal 2 (DG2):} Allow students to easily retrieve correct information and find the relevant document on demand via conversational interaction.

\subsubsection{\textbf{Importance of Interpersonal Communication}}
\label{sec:interpersonal}
One notable finding from the formative study was that, although professors recognized the value of an organizational memory document in reducing the burden of answering questions, they still preferred to communicate directly with students and respond to questions for educational and mentoring purposes. 
In particular, when the questions concerned core skills they aimed to develop, professors expressed a strong desire to answer them directly, even if the questions were repetitive.

\begin{quote}
    \textit{``I do worry sometimes that some things I do prefer to repeat in person, even if it's painful for me. You know, it takes time, etc., right? Because there's some weights carried with delivering it there and saying the same spiel over and over again. Sometimes it's a good wake-up call for people to realize it's important.''} (P11)
\end{quote}

Furthermore, a few professors expressed concern that a chatbot might reduce mentoring opportunities if it answered all questions, thereby removing the need for students to communicate with their advisors. 
One professor, P3, who opposed the idea of organizational memory in general, explained why professors value direct communication with students when we presented multiple scenarios where a student asked a chatbot a question ``How many hours is a student expected to work per week?''
\begin{quote}
    \textit{``I actually don't think I need it, and I actually think this can be counterproductive, honestly. The reason being that, like for this kind of question, I really see it as an opportunity to have a conversation with a student personally because I want to, I don't think it's the number of hours that matter, but I really want to figure out why the student is asking that question.''} (P3)
\end{quote}

\noindent Similarly, some professors also felt that delegating questions to senior students, even though ``senior students will find it annoying (P1),'' can provide valuable opportunities to build a sense of community and offer senior students mentoring experience.

These findings resonate with prior research, which has shown that organizational memory systems alone are insufficient without the social and contextual practices that make knowledge useful~\cite{ackerman2004organizational,pipek2009infrastructuring}. Education research similarly highlights that repeated questions, while seemingly inefficient, play a key role in mentoring and learning by fostering trust, socio-emotional support, and community building~\cite{kram1985mentoring, roscoe2007understanding,lave1991situated}.

Having organizational memory and fostering interpersonal communication in the lab are two seemingly conflicting goals, emphasizing efficiency and education, respectively. To address the professors' concerns, we design the system to actively promote communication between students and professors as well as among lab members. 
We summarize this design goal as follows:

\vspace{0.5em}
\noindent\textbf{Design Goal 3 (DG3):} Facilitate interpersonal communication and integrate the retrieval and update of organizational memory within the context of communication.

\subsubsection{\textbf{Documentation as a Collective Effort with Barriers}}

Professors expressed that organizational memory should be a collective effort, rather than the sole responsibility of lab directors. 
This view was driven not only by the potential to reduce the effort required to answer questions, but also by the recognition that some students are better positioned to explain certain types of knowledge. 
Such knowledge includes hands-on experiences that professors may not be familiar with or processes that only students go through (e.g., submitting documents for Ph.D. requirements).

\begin{quote}
  \textit{ ``If it's something like, how do I request reimbursement for conference travel? You know, it's not going to be more helpful for me to explain that than if my senior student explains it, so I might delegate something like that''} (P9)
\end{quote}

In addition, those who maintained some form of organizational memory wanted to encourage collaborative editing by giving students a certain level of write access, allowing them to contribute by adding and updating content in the documents.

\begin{quote}
    \textit{`` You know you have the login, go and if there's something wrong, or if your internship was different and you find a new rule, it's a Wiki that they can update to. [...] You're part of this lab. Go update the information if it's out of date. I find that to also be kind of a way to make them read it if they feel like they own it as well.''} (P11)
\end{quote}

\noindent In both cases, they noted that students' \textit{ownership} of the document could be beneficial, both in encouraging them to read it and in fostering a sense of contribution or achievement. However, students expressed reluctance toward documentation work, viewing it as tedious.

\begin{quote}
    \textit{``The documentation part is always tedious. Nobody wants to do that. [...] There are times when you feel like, okay, this could be automated or this could be a dedicated job for an AI agent or something like that.''} (S2)
\end{quote}

\noindent Still, professors wanted to track all changes to ensure that edits made by students were legitimate. Ultimately, professors wished to remain aware of the changes and retain the final decision on whether to approve them.

\begin{quote}
    \textit{``I would probably give the students like suggest mode access where they can annotate the document and add a comment or add a suggested edit, and then I would be the one to actually approve it. It wouldn't become live until I made those approvals. And that way I could maintain the accuracy and quality of the documents, and I could be aware of any changes that are happening because I would have to approve them first.''} (P9)
\end{quote}

\noindent Overall, professors viewed it positively when students developed a sense of ownership and contributed to building organizational memory collectively. However, students expressed reluctance toward such documentation work, highlighting a tension between professors' expectations and students' actual willingness to participate.

Prior work has shown that granting editing privileges and fostering a sense of community ownership can encourage sustained contributions and effective use of shared documents~\cite{wagner2004wiki, forte2005why, morgan2013tea}. At the same time, professors emphasized the importance of tracking all changes and retaining final approval rights to ensure accuracy, reflecting a recurring tension also observed in large-scale collaborative systems. 
In Wikipedia, for instance, the balance between openness and quality control has been identified as a central challenge
~\cite{geiger2010work, arazy2011information}. Building on these insights, our design aims to encourage lab members' participation in curating organizational memory while ensuring that professors can maintain oversight and quality control.

\vspace{0.5em}
\noindent\textbf{Design Goal 4 (DG4):} Support members' participation in curating organizational memory and ensure accuracy.


\begin{figure*}[t]
    \centering
    \includegraphics[width=0.86\textwidth]{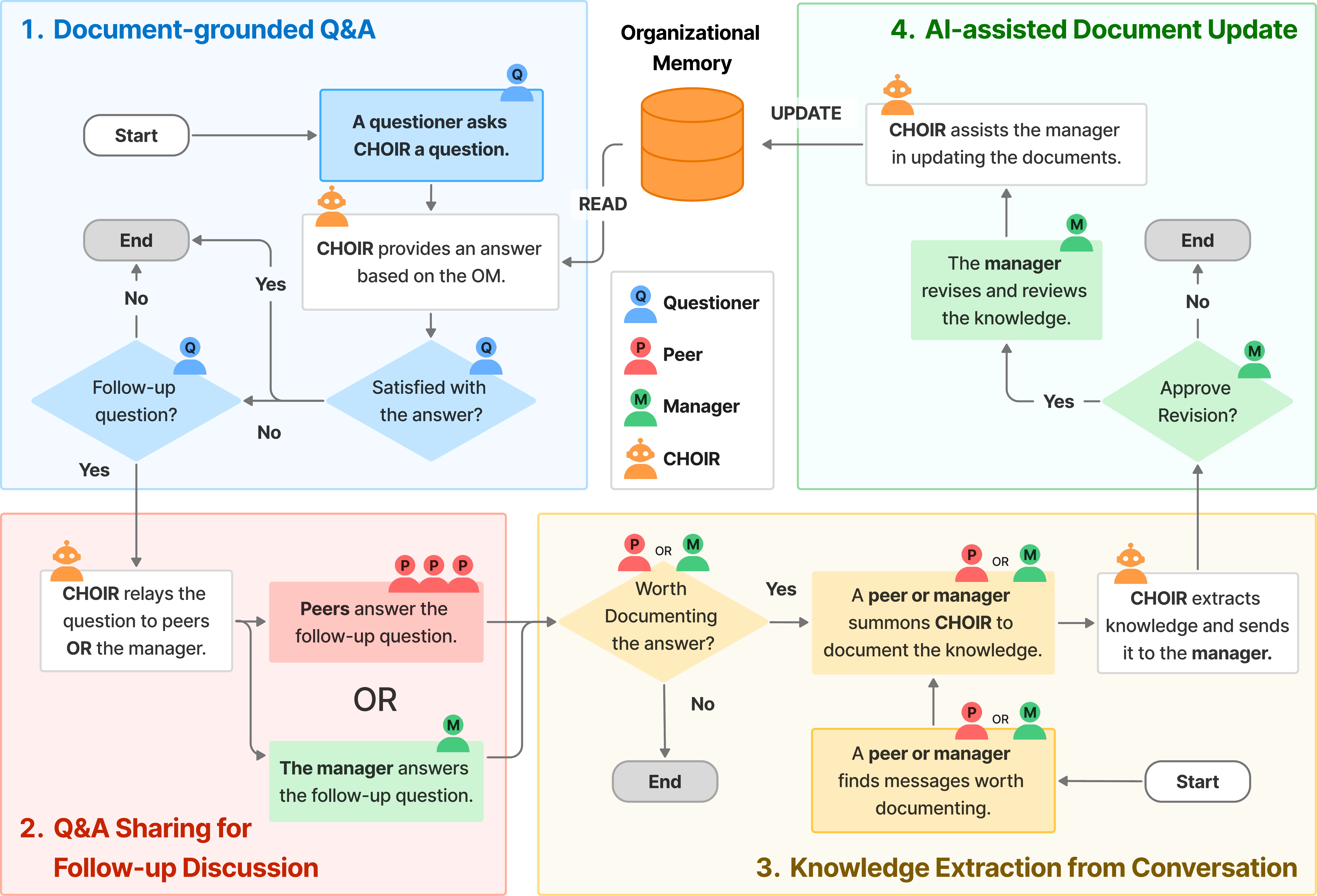}
    \caption{Overview of \sys{}'s workflow. A questioner may ask \sys{} and receive a document-grounded answer, optionally share the Q\&A exchange for broader discussion, and from these interactions, knowledge can be extracted and reviewed by the manager. Approved updates are written back to the organizational memory repository, which also supplies sources for Q\&A.}
    \Description{Flow diagram with four main components: Document-grounded Q\&A with \sys{} (actor: questioner asking a question); Q\&A Sharing for Follow-up Discussion (actors: questioner sharing the Q\&A exchange, peers and the manager contributing insights); Knowledge Extraction from Conversation (actors: peers or the manager suggesting \sys{} save knowledge, or directly providing new knowledge); and AI-assisted Document Update (actor: the manager reviewing and applying changes). The organizational memory repository is placed centrally, connected by READ arrows for Q\&A retrieval and UPDATE arrows for approved changes. Each decision node is controlled by human actors---questioners, peers, or the manager---and branches exist where no updates are made if actors choose not to proceed.}
    \vspace{-10pt}
    \label{fig:choir_flow}
\end{figure*}

\begin{figure*}[b]
    \centering
    \vspace{-20pt}
    \includegraphics[width=0.92\textwidth]{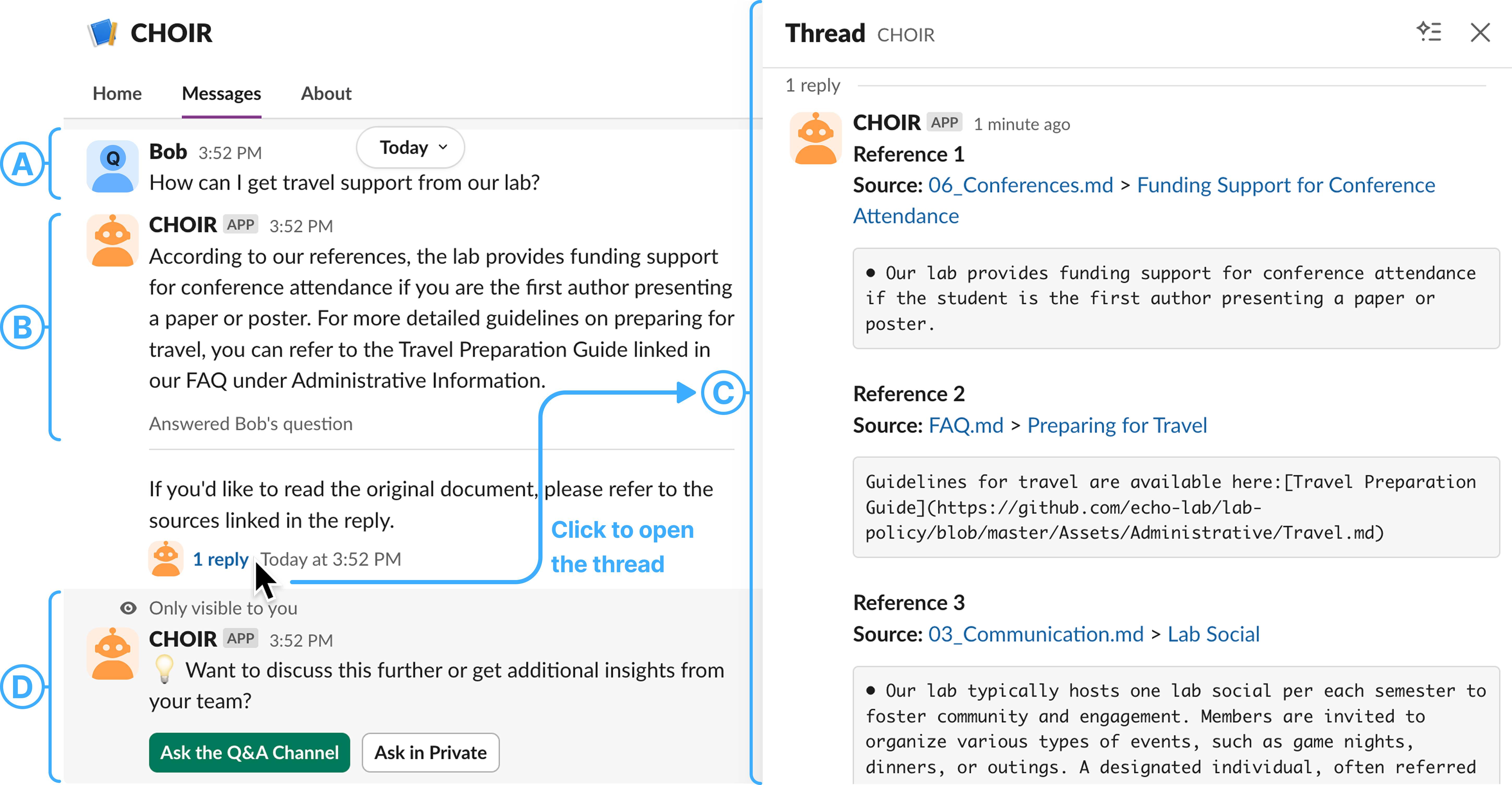} 
    \caption{Document-grounded Q\&A with \sys{} (DM example). Bob (questioner) asks \sys{} a question via direct message, \sys{} returns a document-grounded answer, and the thread is opened to view cited references.}
    \Description{Slack UI in DM mode, with four numbered callouts: (1) member's question submitted via direct message; (2) \sys{}'s grounded answer in the main pane; (3) evidence panel in the thread listing References 1–3 with quoted snippets and source links; (4) share banner, visible only to the questioner, offering two buttons: "Ask the Q\&A Channel" and "Ask in Private." This illustrates the DM case, though questions can also be submitted by mentioning \sys{} in a channel.}
    \label{fig:ui_qna}
\end{figure*}
\section{\sys{}: Chat-based Helper for Organizational Intelligence Repository}\label{sec:system}

\begin{figure*}[t]
  \centering
  \includegraphics[width=0.9\textwidth]{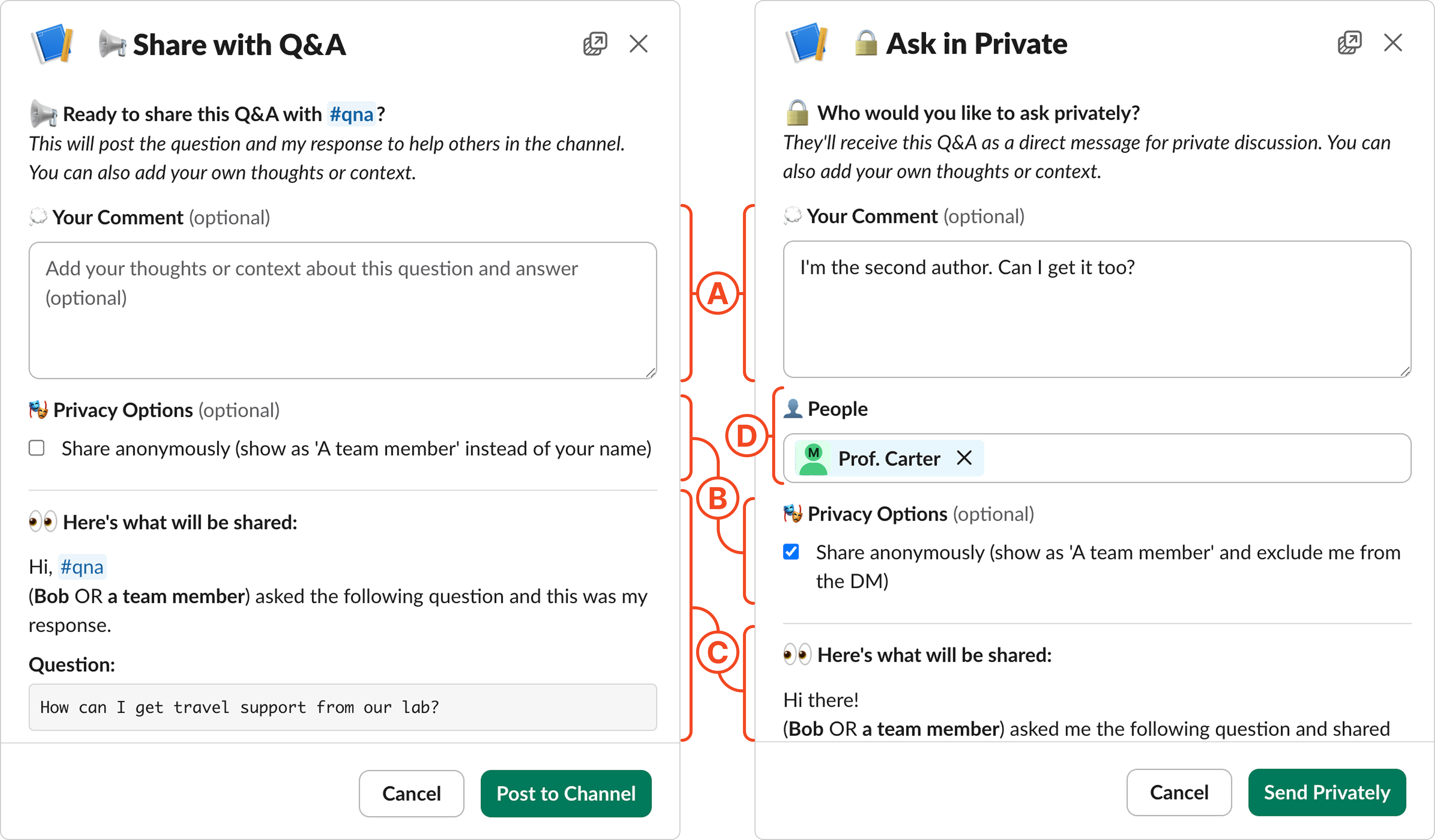}
  \caption{Two share modals in \sys{}. Left: "Share with Q\&A" for posting a Q\&A exchange to the designated channel. Right: "Ask in Private" for sending the same exchange via DM; in this example, the questioner adds a follow-up, the manager is pre-selected in the People picker, and anonymous sharing is enabled.}
\Description{Side-by-side Slack modals. Left ("Share with Q\&A"): (1) optional comment field; (2) anonymous sharing option; (4) preview of the channel post including the Q\&A exchange. Primary action: "Post to Channel." Right ("Ask in Private"): (1) comment field containing an additional follow-up; (3) People picker pre-populated with the manager (editable); (2) anonymous option checked (excludes the questioner from the DM and shows "A team member"); (4) preview of the outgoing DM. This figure illustrates a DM-started case; questions can also be asked by mentioning \sys{} in a channel.}
  \label{fig:ui_share}
\end{figure*}

\begin{figure}[t]
  \centering
  \includegraphics[width=\columnwidth]{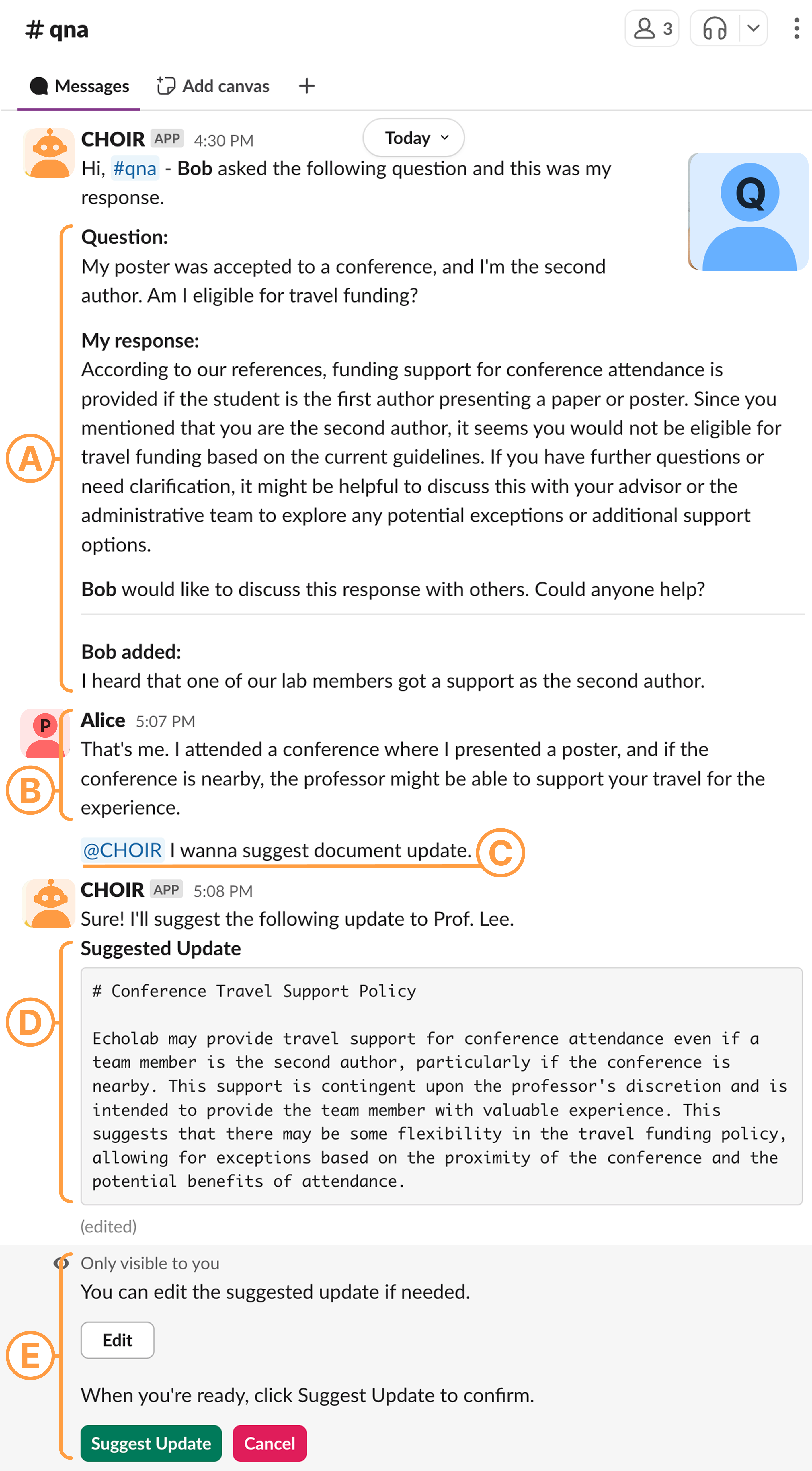}
  \caption{Example of knowledge extraction after a Q\&A is shared. The Q\&A exchange has been posted to the designated channel through the Share with Q\&A modal. A peer (Alice) adds a follow-up comment and then mentions \sys{} to request an update, prompting \sys{} to extract documentable knowledge from the conversation.}
  \Description{Left: the shared Q\&A appears in the qna channel, showing the question, \sys{}'s answer, the questioner's comment, and the profile image (or "A member" if anonymous). Below, Alice posts a follow-up comment. Right: Alice mentions \sys{} to suggest an update, and \sys{} responds with a draft "Suggested Update" block that can be edited and submitted to the manager.}
  \label{fig:ui_extract}
\end{figure}

\begin{figure}[t]
  \centering
  \includegraphics[width=\columnwidth]{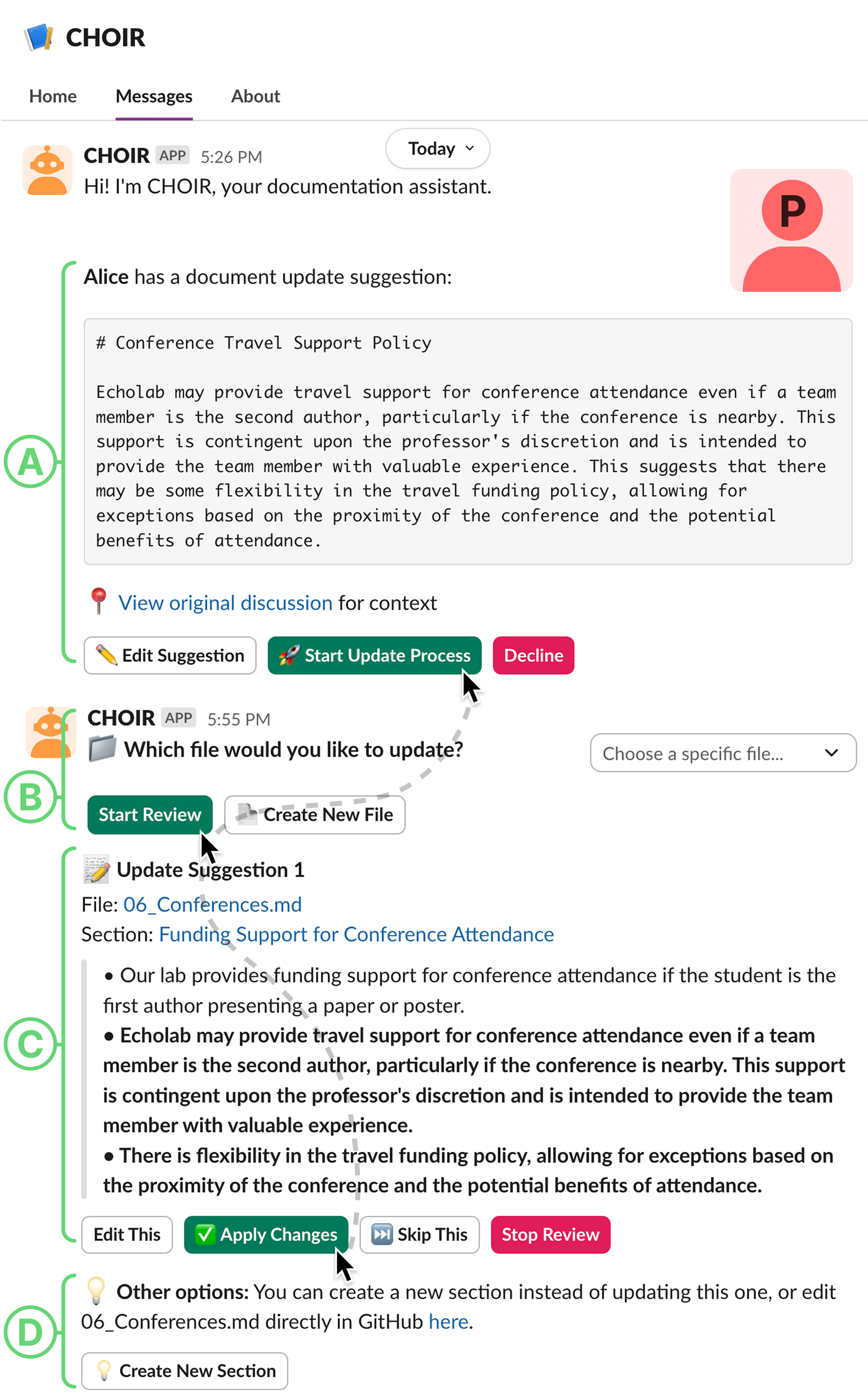}
  \caption{Manager-facing document update workflow in \sys{}. Left: a member's suggestion (from Alice) arrives as a Slack DM with attribution and a link to the original discussion. Right: after starting the update process, the manager reviews the proposed change in context and decides how to incorporate it into the documentation.}
  \Description{Two-panel screenshot. Left: proposal intake card with author attribution, suggestion text, and controls (Edit Suggestion, Start Update Process, Decline), followed by a file picker and Create New File. Right: diff-like suggestion view with bold and strikethrough to indicate edits within a named file and section; action buttons include Edit, Apply, Skip, Stop, plus a Create New Section option and a link to edit in GitHub.}
  \label{fig:ui_update}
\end{figure}

Grounded in our formative study, we designed \sys{} to support students to obtain answers directly within Slack and to transform everyday conversations into a maintained organizational memory. 
\sys{}'s primary objective of sustaining the growth of—and easy access to—organizational memory translates into several subgoals derived from our formative study. These subgoals include streamlining document updates (DG1), enabling easy and accurate information retrieval (DG2),  facilitating communication among members (DG3), and supporting their participation in managing organizational memory (DG4). We accomplish these goals by embedding \sys{} as a chatbot within the lab's communication platform, i.e., Slack.
\sys{} integrates four features for the subgoals: \textbf{Document-grounded Q\&A}, \textbf{Q\&A Sharing for Follow-up Discussion}, \textbf{Knowledge Extraction from Conversation}, and \textbf{AI-assisted Document Update}. These features reflect the design goals identified in the formative study.

When a message is sent to \sys{}—either through a direct message or by mentioning its handle (i.e., \textsf{\textcolor{cyan}{@CHOIR}}) in any channel that it is part of, \sys{} processes it as either a question or an update request, corresponding to the two start nodes shown in \autoref{fig:choir_flow}. 
The four features in \sys{} can support a feedback loop of how a question asked by a student can eventually update the lab handbook.
For example, the member who posed a question may decide to share the question with their peers, and the peers can answer the question, suggesting updating the document. Finally, the manager will decide whether to apply updates. 
The flow may also begin outside of Q\&A, when a member or the manager directly provides new knowledge to \sys{}.
\sys{}'s four features will be detailed in the remainder of this section, showing how each aligns with a design goal (DG1–DG4).

\subsection{Document-grounded Q\&A with \sys{}}

\sys{} allows lab members to ask a question and retrieve information from the documentation in any Slack channel. When a question is submitted, \sys{} answers by composing a response grounded in those sources (\textbf{DG2}).
To avoid hallucination, \sys{} includes an \emph{answerability guard}: when the repository does not provide sufficient coverage, \sys{} abstains from inventing an answer, surfaces related passages with citations, and clarifies what information is available (e.g., related sections or partial guidance). This makes documentation gaps visible for further update, and keeps replies aligned with current documents.

Each Q\&A exchange is posted with grounding and provenance so that the questioner can quickly verify sources. \autoref{fig:ui_qna} highlights the UI elements: the member's question (\bluecirclewrap{A}), \sys{}'s answer (\bluecirclewrap{B}), an evidence panel listing the specific chunks used with anchored links (\bluecirclewrap{C}), and a share banner (\bluecirclewrap{D}) for further discussion. These references encourage the questioner to check out the original documents, thereby getting familiar with the organizational memory.

\subsection{Q\&A Sharing for Follow-up Discussion}

After \sys{} provides an answer, the questioner can share the Q\&A exchange with others to facilitate communication among members (\textbf{DG3}). The share banner (\bluecirclewrap{D} in \autoref{fig:ui_qna}) offers two options: posting to a public Q\&A channel or sending via direct message to selected members. Questioners can configure additional options through one of two modal windows shown in Figure~\ref{fig:ui_share}.  The \emph{Your Comment} field (\redcirclewrap{A} in Figure~\ref{fig:ui_share}) allows the questioner to add context or a follow-up prompt. 

For the ``Ask in Private'' option, the question asker can choose lab members that they will share the question with using a \emph{People picker} (\redcirclewrap{D} in Figure~\ref{fig:ui_share}), which is pre-populated with the manager by default but can be edited to include any Slack members (e.g., senior Ph.D. students).
\emph{Privacy Options} (\redcirclewrap{B}) enable anonymous sharing, in which case \sys{} will replace the questioner's name with ``A team member'' when the message is shared in a Q\&A channel to stay anonymous. In addition, when anonymity is enabled in the private-sharing modal, the questioner is also excluded from the group chat conversation to preserve anonymity among recipients. The preview panel (\redcirclewrap{C}) shows exactly how the question will be shared, either as a channel post or as an outgoing DM. 

The shared Q\&A exchange includes the question, \sys{}'s answer, the questioner's comment, and profile image as shown in \autoref{fig:ui_extract} (\orangecirclewrap{A}), where other members can add follow-up comments (\orangecirclewrap{B}). When shared anonymously in a private DM, \sys{} forwards replies to the original questioner without revealing their identity. When such discussions lead to new knowledge, they provide the entry point for \emph{Knowledge Extraction from Conversation}.

\subsection{Knowledge Extraction from Conversation}

\sys{} supports turning any Slack messages into reusable knowledge (\textbf{DG1}), allowing members to participate in organizational memory maintenance (\textbf{DG4}). Follow-up discussions after Q\&A sharing often contain answers worth documenting. When any lab member sees new knowledge worth keeping, they can summon \sys{} by mentioning \textsf{\textcolor{cyan}{@CHOIR}} (\orangecirclewrap{C} in \autoref{fig:ui_extract}). \sys{} then examines the conversation, identifies \emph{documentable} content, and drafts a suggested update (\orangecirclewrap{D}). The requester can edit this draft and press \emph{Suggest Update} to forward it to the manager (\orangecirclewrap{E}). \sys{} can be summoned anywhere in Slack—for example, if a professor shares a writing tip while giving feedback to a student, the professor can summon \sys{} to document that knowledge, as represented by the new start node in the bottom right corner of Figure~\ref{fig:choir_flow}.

\subsection{AI-assisted Document Update}

This feature of AI-assisted update supports \textbf{DG1} and \textbf{DG4} in multiple ways. Once any member suggests an update from the Knowledge Extraction feature by pressing the ``Suggest Update'' button (\orangecirclewrap{E} in \autoref{fig:ui_extract}), \sys{} delivers it to the manager as a Slack DM message (\greencirclewrap{A} in \autoref{fig:ui_update}) from \sys{}. The message presents the proposed update together with the contributor's name and profile image, making the source and context explicit. At this stage, the manager can edit the suggestion, start the update process, or decline it, and these choices are visible not only in the DM but also as a message in the original channel so that other members are kept updated on how the update request is in progress. This final approval step, which the manager can only handle, is designed to help maintain the accuracy of documented information, which is relevant to \textbf{DG4}.

Once the manager starts the update process by pressing the ``Start Update Process'' button, \sys{} will send follow-up messages one step at a time, guiding the manager to access which file to update, and how. First, the manager can select a target file from the repository or create a new file if the knowledge does not fit into an existing one (\greencirclewrap{B} in \autoref{fig:ui_update}). If not selected, \sys{} will use the most relevant file depending on the similarity of the suggested update and document content. \sys{} then shows which file is selected and the proposed change in a diff-style message with inline emphasis on additions and removals. For each suggestion, the manager may edit the content directly, apply the change to the document, skip it to show the next relevant document --- as there can be multiple places to update in a document, or stop the review entirely (\greencirclewrap{C}). If the knowledge does not fit into an existing section, the interface provides an option to create a new section through a dedicated modal (\greencirclewrap{D}). Applied updates—whether to an existing file, a new file, or a new section—are written back to the repository, and \sys{} posts a message to the original channel acknowledging that the update has been finished. 

This feature implements \textbf{DG1} by automating key steps: recommending where updates are needed, proposing content, and committing changes. Managers can update documents directly in Slack without switching tools. \sys{} preserves managerial autonomy by allowing edits to proposed content or, for more complex tasks such as changing headings or reorganizing structure, enabling direct changes on GitHub via provided hyperlinks (``\textcolor{cyan}{here}'' in \greencirclewrap{D} in \autoref{fig:ui_update}).

\subsection{System Implementation}

\sys{} uses Slack as the front-end client and is built with the Slack Bolt\footnote{https://api.slack.com/bolt} framework. The backend connects to the lab's GitHub repository through the GitHub REST API\footnote{https://docs.github.com/en/rest}, which serves as the persistent store for documentation in Markdown format. This stack allows Slack to serve as the entry point for everyday interactions, while GitHub provides version control for organizational memory.

For Q\&A, \sys{} maintains an embedding index over the documentation repository and performs retrieval-augmented generation conditioned on the user's question and the retrieved text. We use GPT-4o (temperature=0, max\_tokens=1000) for text generation and text-embedding-3-small for document embeddings via the OpenAI API. Input handling is implemented through prompting: the LLM is instructed to decide whether an incoming message should be treated as a Q\&A query or an update request. Responses to questions, knowledge extraction from conversations, and documentation update drafts are all generated via the OpenAI API. To protect privacy, user names are replaced with pseudonyms before any message is sent to the API. The complete system prompts are provided in \autoref{sec:appendix}.

\section{Field Study}

To understand how lab members engage with and experience a chatbot-driven organizational memory system (RQ2), we conducted a month-long field deployment of \sys{} in the Slack workspaces of four university research labs, including the authors' own. 

\subsection{Study Design}

To explore these questions, we conducted an exploratory field deployment study that combined analysis of system usage logs with user interviews~\cite{creswell2017designing}, enabling us to capture both behavioral patterns and the reasoning behind user actions. We deployed \sys{} in active research laboratories to observe usage and experiences within participants' real work environments~\cite{rogers2011interaction}. This field study approach reflects the often underestimated difficulty of evaluating CSCW applications: organizational systems must be assessed within authentic work contexts, as in-lab experiments that attempt to simulate ecologically valid environments cannot reproduce the social dynamics and group interactions that unfold over weeks and months~\cite{grudin1988why}.

Our evaluation focuses on understanding how CHOIR mediates organizational knowledge management practices, thereby demonstrating the feasibility of \sys{} and the benefits observed through user behaviors and interview responses. Field studies of organizational memory systems have long shown how such systems augment collective memory within real organizational contexts and workflows~\cite{ackerman1998augmenting}.  While field deployments cannot establish causal relationships due to the challenges of creating controlled baselines, they enable a holistic understanding via a \textit{descriptive} approach~\cite{lazar2017research} that characterizes how CHOIR mediates organizational communication and memory practices in situ. Many social computing studies that involve a group of users, an organization, or a community, similarly choose to evaluate systems through deploying the system in the wild and takes the descriptive approach~\cite{mahar2018squadbox, 10.1145/1822258.1822283, 10.1145/3432912, bali2023nooks, 10.1145/3173574.3173769, 10.1145/2145204.2145249, 10.1145/3025453.3025780}. In studying sociotechnical interventions in organizational contexts, authentic workplace dynamics provide crucial insights into user engagement and experience with collaborative knowledge management systems.

\begin{table}[t]
   \centering
   \caption{Participants in the field study: each lab's director and members, including newcomer status. Lab age is shown in parentheses below each lab number.}
   \Description{A table showing 21 participants across 4 research labs (L1-L4) with 4 columns: Lab number with age, Department and research area, Participant label with gender, and Status. L1 is a 4-year-old Software Engineering lab with 1 director (D1, male) and 4 members (1 new female member, 3 existing members). L2 is a 4-year-old HCI lab with 1 director (D2, female) and 5 members (2 new male members, 3 existing members). L3 is a 4-year-old Physical Ergonomics lab with 1 director (D3, female) and 5 members (2 new members, 3 existing members). L4 is a 7-year-old HCI lab with 1 director who is also an author (D4, male) and 4 existing members. The table shows a mix of new and existing members across all labs, with varied gender representation.}
   \label{tab:participants}
   \small
   \resizebox{\columnwidth}{!}{%
   \begin{tabular}{llll}
   \toprule
   \textbf{\shortstack[l]{Lab\\[1pt](years)}} & \textbf{\shortstack[l]{Department\\[1pt](Research Area)}} & \textbf{\shortstack[l]{Label\\[1pt](Gender)}} & \textbf{Status} \\
   \midrule
   L1 & Computer Science & D1 (M) & Director \\
   (4 yrs) &   (Software Engineering) & S1-1 (F) & New student \\
      &  & S1-2 (F) & Existing student \\
      &  & S1-3 (M) & Existing student \\
      &  & S1-4 (M) & Existing student \\
   \midrule
   L2 & Computer Science & D2 (F) & Director \\
   (4 yrs) &    (HCI) & S2-1 (M) & New student \\
      &  & S2-2 (M) & New student \\
      &  & S2-3 (F) & Existing student \\
      &  & S2-4 (M) & Existing student \\
      &  & S2-5 (F) & Existing student \\
   \midrule
   L3 & Industrial Engineering & D3 (F) & Director \\
   (4 yrs) &     (Physical Ergonomics) & S3-1 (M) & New student \\
      &  & S3-2 (F) & New student \\
      &  & S3-3 (F) & Existing student \\
      &  & S3-4 (F) & Existing student \\
      &  & S3-5 (M) & Existing student \\
   \midrule
   L4 & Computer Science & D4 (M) & Director (Author) \\
   (7 yrs) &   (HCI) & S4-1 (M) & Existing student \\
      &  & S4-2 (M) & Existing student \\
      &  & S4-3 (M) & Existing student \\
      &  & S4-4 (F) & Existing student \\
   \bottomrule
   \end{tabular}%
   }
\end{table}

\subsection{Participants and Recruitment}

We recruited university research lab directors through our social media and from formative study participants who expressed interest in the field study. Four research laboratories participated in the user study, comprising 22 participants: four lab directors and 18 students. Upon confirmation from each director, we advertised the study in their Slack workspace and recruited students from the lab, endorsed by the lab director. 
We refer to each unit and participant based on the labels as specified in \autoref{tab:participants}.
The one-month deployment was conducted during the Summer of 2025, a transitional academic period when each lab had remote incoming graduate students joining. This timing allowed us to observe how CHOIR supports early-stage onboarding and knowledge exchange when new members have many initial questions. 

All participants provided informed consent for interviews, system usage monitoring, and interaction log collection. The study received IRB approval (IRB 25-247). As compensation, lab directors received \$40 and students received \$20 upon completing the interviews.
One participating lab (L4) was led by the corresponding author, who served as the lab director. 
\newadd{In addition to easing recruitment, the study was motivated by the director’s experience with manually curating organizational memory and observing the limited use of these materials in practice.} 
To mitigate any potential pressure that L4 participants might feel due to power dynamics or a desire to be favorable toward their peers, the director limited their role to answering questions and making changes to documents and did not otherwise intervene in the study procedures. Similarly, the first and second authors, who are members of L4, did not use \sys{} in this field deployment at all. All L4 participants voluntarily consented to the study independently through the standard procedure of advertisement, recruitment, and onboarding sessions, moderated by the first and second authors. 

\subsection{Deployment Process}

The field deployment occurred in two phases: onboarding and active deployment.

\subsubsection{Phase 1: Document Collection and Onboarding Sessions}
Prior to the deployment period, we collected existing lab documents and conducted structured onboarding sessions with each participating lab.

\noindent\textit{Document Collection:} Before the lab director onboarding sessions, we asked professors to share existing lab documents or materials they wished to integrate with \sys{}. We specifically requested only documents that could be made publicly available. The participating labs exhibited diverse documentation practices. D1 and D3 had no existing organizational memory systems and primarily relied on informal communication channels such as Slack messages, verbal exchanges, or external resources (e.g., department handbooks). To address this and avoid a cold-start problem, the research team worked with them to create initial seed documents based on available resources (e.g., lab websites and departmental handbooks), ensuring that \sys{} would contain useful content immediately upon deployment. In contrast, D2 and D4 already maintained web-based documentation, including a lab webpage and a GitHub repository containing the director's advice, mentoring plan, and lab policies. However, during the study period, they voluntarily used \sys{} rather than their existing systems. The research team, with the directors' approval and guidance, converted all provided materials into markdown files, uploaded them to a public GitHub repository for each lab, and granted lab directors write access and students read access.

\noindent\textit{Lab Director Onboarding Sessions } (45–60 minutes) were conducted via Zoom with lab directors only. These sessions covered \sys{}’s knowledge management capabilities, administrative features, the GitHub repository structure, and manager approval workflows for document updates.

\noindent\textit{Student Onboarding Sessions } (45–60 minutes) were also conducted via Zoom to accommodate as many participants as possible. These sessions focused on \sys{}’s question-answering functionality, conversation facilitation features, and integration with the existing Slack workspace. 
Both onboarding sessions included: (1) a system demonstration with examples drawn from typical lab environments, (2) hands-on practice with key features, and (3) a brief semi-structured interview.

\subsubsection{Phase 2: Active Deployment}
Following onboarding, \sys{} was deployed as a Slack App within each lab’s existing workspace. 
Participants were registered as authorized users with two distinct roles: lab directors served as managers with document approval access, while students operated as \sys{} users.
\newadd{No specific usage requirements were imposed on participants, while awareness of \sys{}'s availability was maintained by having the research team periodically ask questions to \sys{} in shared channels.}
Participant privacy was safeguarded through strict data collection protocols and automatic anonymization processes of personally identifiable information in compliance with FERPA requirements. When eligible messages were sent to external language models (OpenAI API), all usernames were replaced with pseudonyms, and responses were de-anonymized before delivery to users.

\subsection{Data Collection}

We collected two types of data during the one-month field deployment (Summer 2025). 

\subsubsection{System Interaction Logs}
We collected interaction data within the \sys{} application, including message exchanges with \sys{}, feature usage patterns, button clicks, and document update activities recorded both in \sys{} and on GitHub. To protect participant privacy, messages were transferred to our servers or the OpenAI API only in three cases: (1) direct messages sent to \sys{}, (2) messages that mentioned \sys{}, and (3) the ten most recent messages when \sys{} was mentioned for knowledge extraction. 
We tracked usage metrics such as frequency and timing of \sys{} interactions per user, types of questions asked, conversation thread lengths, manager approval rates for suggested updates, and user feedback actions.

\subsubsection{Log-Informed Semi-Structured Interviews}

Following the deployment period, we conducted semi-structured interviews with each lab director and with students who had used \sys{} at least once. Rather than treating logs and interviews as separate data sources, we used individual usage patterns from the logs as prompts during interviews, enabling participants to explain their reasoning and experiences. This approach allowed us to understand not only what users did, but also why they made particular choices and how they experienced the system.

\subsection{Data Analysis}

We analyzed system interaction logs to first generate descriptive statistics of usage patterns of each lab and identify individual usage patterns, including feature adoption, engagement trends, and workflow progress rates. 
These patterns also served as the foundation for personalized interview discussions, allowing us to explore the reasoning behind specific user behaviors.

Transcripts from post-deployment log-informed interviews were analyzed using reflexive thematic analysis~\cite{braun2019reflecting}. Two researchers independently coded a subset of transcripts to establish themes, then collaboratively developed a coding framework applied to all interviews. Key themes focused on how participants used \sys{} for organizational memory, the barriers they faced, and their suggestions for improvement.

\section{Results}

Overall, \sys{} supported knowledge retrieval and updates to organizational memory during the study period, while also revealing nuanced challenges and opportunities for chatbot-mediated knowledge management systems, which we present below.

\subsection{Q\&A Interactions with \sys{}}

\subsubsection{``I Consider \sys{} as a Senior in the Lab}''

During the one-month field study, participants submitted a total of 107 questions to \sys{} through both direct messages and public channels. 
Of these queries, 45 (42\%) were answered using the connected lab documentation, while \sys{} could not answer the remaining questions and revealed gaps in the existing knowledge base. 
Participants valued \sys{}'s effectiveness and reliability, with S2-1 noting that ``\textit{it definitely improved the speed of retrieving the information compared to me reading through the document or searching online.}'' 
S1-2 appreciated its accuracy with reference: ``\textit{if the information is already in the handbook, it's pretty straightforward to get those information, and I like that it provided a reference [...] That helped me to get the accurate part information.}''
D1 appreciated the quality of responses: ``\textit{I think in general the answers were pretty good. When the response or when the like data was there, further details were there.}''

Beyond its functional utility, deploying \sys{} in the social context of Slack led participants to perceive and interact with it as if it occupied familiar social roles. S1-3 conceptualized \sys{} as a knowledgeable lab member: ``\textit{I can go to the \sys{} to ask him, confirm the question[...] I consider \sys{} as a senior in the lab.}'' This perspective was shared by S4-2, who viewed \sys{} ``\textit{more like a librarian. A library librarian tells you what's in the library and not if it's not there.}'' 
These metaphors reflected how participants made sense of \sys{}'s role within their lab ecosystem, framing its consistent availability and non-judgmental responses in terms that helped them integrate AI assistance into their daily lab interactions. When prompted, all lab members expressed a desire to keep \sys{}, and the app remains active at the time of submission.

\begin{figure}
  \centering
  \includegraphics[width=\columnwidth]{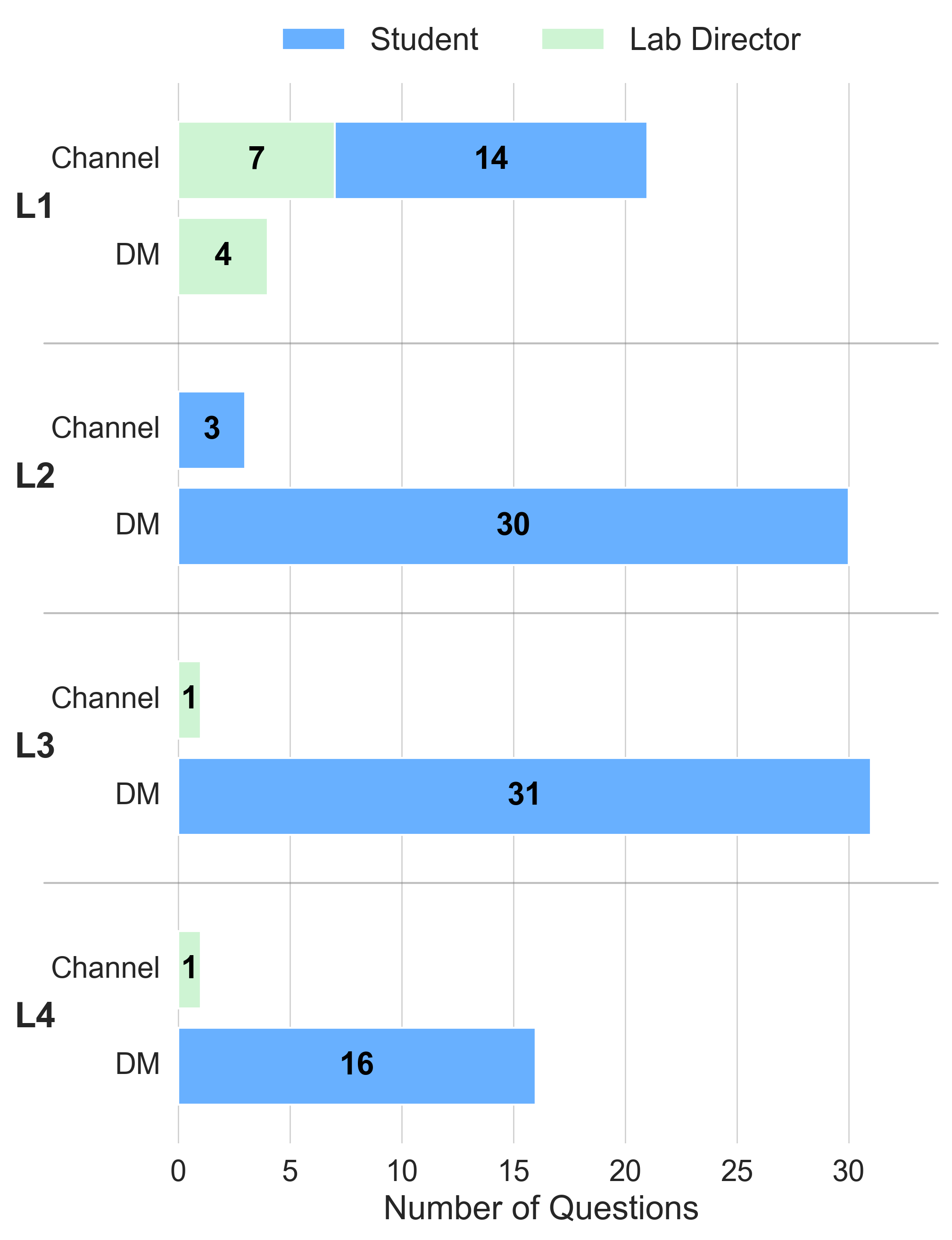}
  \caption{Distribution of questions asked to \sys{} across four research labs during the one-month field study. The graph shows the number of questions submitted through public channels versus direct messages (DM), broken down by students (blue) and lab directors (green). L1 shows a distinctive pattern with predominantly public channel usage, while L2, L3, and L4 primarily used private DMs for Q\&A interactions.}
  \Description{A horizontal bar chart showing question counts across four labs. Each lab has two bars representing Channel and DM usage. L1 shows 14 student questions and 7 director questions in the channel, with 4 director DM questions. L2 shows 3 student questions in the channel and 30 student questions in DM. L3 shows 1 director channel question and 31 student DM questions. L4 shows 1 student public question and 16 student DM questions. The chart uses blue bars for students and green bars for lab directors, with directors shown on the left side of each stacked bar.}
  \label{fig:qa_count}
\end{figure}

\subsubsection{How Lab Culture Shapes Question Asking}
\label{sec:publicqna}
One notable pattern that we found from L1, as shown in \autoref{fig:qa_count}, was that most of the questioners asked \sys{} in a public channel instead of via DM, contrasting sharply with the other three labs. L1 students asked \newadd{14} questions publicly, \newadd{with none through private messages}. Students' motivation to ask public questions stemmed from altruistic and supportive attitudes to help others. S1-3 explained: ``\textit{I think it can share to everybody about the answer[...] everybody can just check the Slack and see.}'' This visibility allowed students who had similar questions to benefit from others' queries without needing to ask again. Some students also used \sys{} to support their lab. S1-4 stated, ``\textit{I mostly used it to help [D1]. Add stuff that wasn't already there.}'' The public nature of these interactions enabled collective awareness and knowledge sharing and also served as a catalyst for documentation improvements, with multiple students explicitly using \sys{} to identify gaps in the handbook.

This distinctive use of public Q\&A perhaps stemmed from their altruistic and supportive lab culture. When D1 was asked to speculate why this pattern showed, he mentioned ``\textit{I think all the students get along pretty well[...] more comfortable with public questions and not being afraid to ask questions.}'' This comfort level was echoed by students, with S1-2 noting, ``\textit{Our lab atmosphere is pretty good. We know each other well, so I don't feel those kinds of questions cannot be asked publicly.}'' This mutual trust and familiarity created an environment where both directors and students felt comfortable sharing their uncertainties publicly, fostering a collaborative learning atmosphere. This collaborative approach resulted in D1 making more frequent updates than any other labs (\autoref{tab:update_patterns}), as questions surfaced with missing information that needed to be added.

In sharp contrast to L1, \autoref{fig:qa_count} shows that L2, L3, and L4 predominantly used private messages. These participants sought a safe space to ask questions while avoiding the potential negative impression they may have from the question. S2-2 expressed, ``\textit{I'm just afraid that my question could be like, could be dumb questions. So I just asked it first through DM because it's more private.}'' 
Participants also avoided public channels to prevent burdening others with potentially trivial questions, with S4-2 noting: ``\textit{when I don't want to bother others,}'' This sentiment was shared by lab directors.

\subsubsection{Directors' Personal Assistant to Convey Knowledge}

While we did not expect lab directors to ask \sys{} questions beyond testing, lab directors repurposed \sys{} to communicate information to students. Directors Q\&A to \sys{} showed different approaches to alternative usage patterns. D1 used public Q\&A primarily as an acknowledgment about documentation updates: ``\textit{after I updated and asked the question, see if the response is correct. Also, maybe kind of to highlight to whoever asked the question that it is updated now.}'' This approach transformed \sys{} into a broadcasting system, allowing D1 to signal to students that their previously unanswered questions had been addressed through documentation improvements. Similarly, D3 asked \sys{} a question once in a public channel to proactively share important information by asking questions such as ``Can you tell how CITI training can be completed?'' immediately after asking new students to take the CITI Training, to give them detailed instructions on which courses they needed to take. 
These cases demonstrate how lab directors leverage \sys{} as an assistant to acknowledge updates or disseminate information rather than simply seeking answers.

\subsection{Knowledge Sharing Through Q\&A}
\label{subsec:share_qa}

\subsubsection{Q\&A Sharing for Collecting Answers as well as Questions}

\sys{}'s Share Q\&A feature enabled users to share their questions and \sys{}'s responses with others, either in the designated Q\&A channel or through private messages. Of the 95 questions asked by students, only 11 were shared (L1: 1, L2: 4, L3: 3, and L4: 3 with 1 shared privately with a director).
Students appreciated the Sharing Q\&A feature for being able to understand the context of someone's question, as S2-3 noted ``\textit{ I also like when it's like sending the questions in the channel, like it's also including what it had. So when we like, try to give our answers, we also have some reference to refer to.}'' In the meantime, L1's distinct pattern of public Q\&A practice introduced in \autoref{sec:publicqna} effectively replaced the Share Q\&A feature as most questions and answers were already visible to all students in L1 through their public interaction style.

Lab directors particularly appreciated the centralized Q\&A channel for its organizational benefits. D2 valued the dedicated space for continuous interaction: ``\textit{I like the Q\&A channel. So it's like those that's for people to ask questions and I can answer them. I think I would like to keep like Q\&A channel [...] If people have questions, they can ask them there.}'' D3 emphasized the channel's role in managing communication flow: ``\textit{I think having this channel can kind of centralize all those messages in one place, so I like it.}'' While Q\&A sharing was primarily designed for getting answers for unanswered questions, it supported DG3 in a way that we did not anticipate. 
Overall, deploying \sys{} in these labs encouraged and fostered the culture of sharing knowledge through Q\&A using various channels: plan Slack communication, asking \sys{} in a public channel, and sharing the Q\&A feature, supporting our goal of facilitating interpersonal communication (DG3).

\subsubsection{Having a Safe Space to Ask and Discuss}

Beyond these patterns, \sys{} provided a safe channel for students to ask questions anonymously, even when the contained anonymity came from the small scale of the participants. Notably, more than half of the shared questions (6/11) were posted anonymously, indicating users' desire for privacy even when seeking broader input, which could have kept them from sharing it if it were not anonymous. 
S3-1 explained their use of anonymity: ``\textit{maybe thinking the question is kind of stupid question[...] if some senior student may think, okay, this question, maybe I'm supposed to know it, maybe it's in the graduate manual, but I just didn't read it. If I ask people going to think I'm stupid and lazy.}'' Participants also cited finance- or budget-related topics as hypothetical examples that might warrant anonymous sharing, with S2-2 noting, ``\textit{sometimes I don't want to reveal myself because I know there are some sensitive information like finance or budget.}'' One such example involved a new Ph.D. student anonymously asking their peers when they would receive their first stipend—an important question that could have been awkward to pose to an advisor, who might not know the answer, and one the student might not feel comfortable asking of any peers. 

Students had an additional motivation to share a question using \sys{} as a way to initiate discussions on topics where people might hold different opinions. Rather than seeking definite answers, participants shared questions to collaboratively explore a topic (e.g., `How many papers are needed to graduate?'), rather than expecting to get a definite answer. S2-2 stated, ``\textit{I just want to see like how others think about the topic,}'' and S4-4 sought ``\textit{some suggestions from different people, their personal suggestions instead of an article.}'' These examples demonstrate how \sys{} created a space for students to exchange perspectives and engage in open-ended conversations that they did not necessarily think should be documented in the organizational memory.

\subsubsection{Question Sharing Complements, not Block, Interpersonal Communication}

Despite the availability of sharing features, the vast majority of questions (86 out of 95) remained unshared, raising questions about both social and practical barriers to collaborative question and knowledge sharing. Instead, students chose to communicate directly with peers instead of \sys{}-mediated sharing. S3-3 explained, ``\textit{the most straight path for me was to just ask D3 in a direct message, rather than using \sys{} as, like a middleman.}'' Similarly, S2-3 reported, ``\textit{I didn't ask it anonymously in the Q\&A channel because I know if there is someone who could answer my question, then it would be a specific senior PhD student, and I would probably just go to them and ask them directly.}'' Overall, the limited activity of Q\&A sharing was not necessarily due to barriers or fear; rather, students often chose traditional modes of communication to obtain answers more directly.

\subsubsection{When Questions Stay Hidden: Barriers to Director Awareness}
\label{sec:hidden}
The limited activity in Sharing Q\&A created a significant awareness gap for lab directors, particularly those in L2 and L3, who had limited visibility into what questions students were asking and which gaps existed in their documentation. During post-study interviews, in general, D2 and D3 wondered if anyone really used \sys{}, although the students were actively asking questions about it behind the scenes, which came to their surprise when we presented the statistics.  D3 reflected on this limitation: ``\textit{I didn't receive any request [to answer questions] directly [...] I also don't know what kind of questions they've asked.}'' While we anticipated that the Sharing Q\&A feature would raise awareness of gaps in the documentation, students’ preference for direct communication limited the lab directors’ view on what students want to know and \sys{} do not know.
D2 recognized the value of knowing student questions: ``\textit{it might be helpful for me to know what questions are asked so that I know what they are confused about [...] if I don't know what they are confused about, then there's nothing I can do to clarify the confusion.}'' This lack of visibility prevented directors from identifying knowledge gaps, which is crucial for understanding student confusion, or updating documentation proactively. In contrast, L1's culture of public Q\&A provided natural visibility into student needs and documentation gaps, enabling more responsive documentation maintenance.

\subsection{Knowledge Extraction and Document Update Practices}
\label{subsec:knowledge_extraction}

\subsubsection{Perceived Value of Document Update via \sys{}}

Participants recognized significant value in collaborative documentation, particularly appreciating the potential for preserving and sharing knowledge within research labs. The organizational value of maintaining documentation was widely acknowledged, with D1, who did not have any organizational memory beforehand, expressing satisfaction: ``\textit{the documents. I'm very satisfied [...] I think it'll continue to live on and continue to be updated even after the study.}'' D2 emphasized the importance of capturing firsthand student experiences: ``\textit{because they have firsthand information [...] if they don't document this, then the next person would have to figure it out by themselves again.}'' 
S1-1 highlighted the cumulative value for future members: ``\textit{the more questions asked and updated, the more useful it becomes for future students.}''

Directors particularly valued the AI-assisted update workflow, which reduced the effort required for documentation maintenance. D3 appreciated \sys{}'s editing suggestions: ``\textit{\sys{} suggested me a sort of edit, which I liked [...] I don't have to spend a lot of time crafting.}'' D2 found the process straightforward, noting, ``\textit{I just click add and I think it asks me to select a document.}'' Directors who either received update requests from students or initiated updates themselves incorporated them into the lab documentation, suggesting the effectiveness of the workflow when participants chose to engage.

\begin{table}[t]
   \centering
   \caption{Documentation update patterns across research labs during the one-month field study, showing commits, files modified, and word changes for both direct GitHub updates and \sys{}-assisted updates.}
   \Description{A table with 5 columns showing documentation update statistics for 4 research labs (L1-L4). Each lab has two rows: Direct Update and CHOIR system updates. L1 shows 19 direct commits with 1153 words added and no CHOIR usage. L2 shows minimal direct updates (1 commit) but 4 CHOIR commits with 261 words added. L3 shows no direct updates and 1 CHOIR commit with 109 words added. L4 shows balanced usage with 7 direct commits and 6 CHOIR commits, totaling 470 words added across both methods.}
   \label{tab:update_patterns}
   \small
   \resizebox{\columnwidth}{!}{%
   \begin{tabular}{llrrr}
   \toprule
   \textbf{Lab} & \textbf{\shortstack[l]{Update\\[1pt]Method}} & \textbf{Commits} & \textbf{Files} & \textbf{Words (+/-)} \\
   \midrule
   L1 & Direct & 19 & 5 & +1153 / -63 \\
      & \sys{} & 0  & 0 & 0 \\
   \midrule
   L2 & Direct & 1  & 1 & +1 / -1 \\
      & \sys{} & 4  & 1 & +261 / -0 \\
   \midrule
   L3 & Direct & 0  & 0 & 0 \\
      & \sys{} & 1  & 1 & +109 / -1 \\
   \midrule
   L4 & Direct & 7  & 2 & +122 / -29 \\
      & \sys{} & 6  & 2 & +348 / -2 \\
   \midrule
   \textbf{Subtotal} & Direct & {\textbf{27}} & {\textbf{8}} & \textbf{+1276 / -93} \\
    & CHOIR & {\textbf{11}} & {\textbf{4}} & \textbf{+718 / -3} \\
   \midrule
   \textbf{Total} & -- & {\textbf{38}} & {\textbf{12}} & \textbf{+1994 / -96} \\
   \bottomrule
   \end{tabular}%
   }
   \end{table}

\subsubsection{Directors' Update Practices and Member Engagement}

During the study, lab directors exhibited varying update patterns, as summarized in \autoref{tab:update_patterns}. 
D3's only knowledge extraction attempt was made to update the document in a direct message with \sys{}. On the other hand, D2 proactively used \sys{}'s knowledge extraction four times, completing four varying types of document updates. 
Two document updates originated from private Q\&A sessions where S2-2 shared questions privately, prompting D2 to provide direct answers and mention \sys{} to trigger knowledge extraction. Another extraction occurred when S2-3 and S2-5 shared their experiences in response to an anonymous question shared in a channel, leading D2 to mention \sys{} to capture students' answers into the document. The fourth extraction involved a more complex collaborative process where D2 initially provided an answer while mentioning senior students to supplement the answer with additional information; when S2-3 subsequently provided the requested information via message, D2 mentioned \sys{} again to extract and document this enhanced response. This variety of cases exemplified how professors can leverage the collaborative and communicative nature of the workspace for documentation updates even when they lack hands-on knowledge in specific areas.

Beyond these expected uses, others also actively updated documents through the document editor available in GitHub. D4, one of the authors, showed a balanced approach, using both direct updates (7 commits) and \sys{}-assisted updates (6 commits) to maintain their documentation.  D1 performed updates outside \sys{} through GitHub after identifying gaps via public Q\&A, resulting in the highest volume of documentation changes with 19 commits across 5 files and 1,153 words added.
When asked why he directly edited, D1 explained, ``\textit{Probably just out of convenience. It was easier. Or maybe I was just more familiar with pulling up the repo and making a change directly.}'' Given that he started with nearly empty documents, he would not have faced the burden of locating specific sections to update, as someone managing a longer document might have. 

\subsubsection{Barriers to Student Contributions in Documentation}

Although some students contributed to the knowledge when their lab director specifically requested, as in D2's case above, students did not actively engage in initiating document updates; only three participants (S2-2, S2-3, S3-2) attempted to use the feature, all through private messages. These three members initiated a total of five updates during the study period. Although we expected students to request updates through the public Q\&A channel, they primarily discovered documentation gaps during private Q\&A with \sys{}, providing corrections based on their own knowledge before proposing updates to directors. For example, S3-2 noted, ``\textit{I was asking about which courses to take, and \sys{} talked about the two required courses, but the course code was incorrect [...] And so I just updated it.}'' This pattern suggests the feasibility of \sys{} in enabling students to contribute to documentation.

Despite recognizing its value, students’ participation in documenting updates remained limited due to several concerns. Knowledge-related concerns emerged as the primary obstacle. A lack of confidence in providing correct and accurate answers was a central issue, with S3-3 worrying about the permanence of contributions: ``\textit{if I answered that question that would stay in the database forever [...] if it was wrong, then it would mislead all the students.}'' S2-3 expressed a similar hesitation: ``\textit{sometimes I'm not sure about the answer[...] the wrong message would be there for some time and it might mislead.}''

Students’ limited understanding also prevented them from generalizing personal experiences into knowledge useful for everyone, posing another challenge. Participants reported that not all knowledge was suitable for general documentation. S3-4 explained, ``\textit{especially regarding the funding sources, some things are not generic, they can be very specific to someone.}'' 
At the stage of design, we believed that DG4 would be supported by having professors act as gatekeepers of how knowledge is documented in organizational memory. However, our findings indicate that students' psychological barriers can be a larger issue that sheds lights on the need to foster a culture where documentation is viewed as a collective and welcoming practice rather than an individual risk.

\subsection{Participant Suggestions for Improvement}

Throughout the study, participants identified limitations in \sys{}'s capabilities and offered suggestions for improvement based on their experiences. These recommendations revealed both immediate needs and longer-term visions for AI-mediated knowledge management in research labs.

\subsubsection{Manager Awareness and Communication}

For lab directors, the predominance of private Q\&A created a particularly problematic awareness gap as stated in \autoref{sec:hidden}.
To address this critical gap, participants suggested several solutions for improving manager awareness while preserving student privacy. Participants suggested that students' questions be aggregated for periodic high-level summaries. S2-4 proposed to ``\textit{take all of them, strip the metadata, and then have a nice summary of the key points and send that as a weekly update to the professor.}'' 
The need for users' option to opt out from being reported was also suggested, with S3-4 emphasizing that ``\textit{there should be an option. If we want to send the question to the advisor or not.}'' These suggestions reflected a nuanced reluctnace that they have for contributing to the organizaionl memory and preserving student autonomy and privacy.

\subsubsection{Expanding Knowledge Base Access}

Participants expressed frustration with \sys{}'s limitation to internal documentation—a constraint stemming less from the system itself than from the limited content available in the repository—and suggested multiple avenues for expanding its knowledge sources. For external resources, D3 noted that ``\textit{it would be better that we can search relevant information from other sources,}'' while S3-3 proposed that ``\textit{if it can access some websites or other miscellaneous web pages or some forums, maybe it would be more helpful.}'' Integration with internal resources was considered equally important, with S2-5 expressing strong interest in ``\textit{access to the Google Docs. I think in our group, the advisor mentioned everything in the Google docs.}''
Beyond documents, participants envisioned \sys{} as a bridge across labs and expertise sources. Cross-lab knowledge sharing emerged as particularly valuable, with D2 expressing interest in seeing ``\textit{what questions students ask[...] if the answers are shareable, I would like to see their answers too.}'' Similarly, S2-5 suggested leveraging other labs' experiences: ``\textit{if there are datas of other lab policies[...] \sys{} could give suggestions on what is missing in our lab policy and elicit discussions.}''
These suggestions point toward a more interconnected knowledge ecosystem—one that bridges internal and external resources, integrates documents and people, and extends across different research groups.

\subsubsection{Context-Aware and Personalized Support}

Participants imagined a system that could adapt to different contexts and individual needs rather than providing one-size-fits-all responses. For situation-specific guidance, S3-3 suggested differentiated support: \``textit{if there can be a process where there are some situations created, for first-year or second-year PhD students, then maybe there can be some other ways to update that information.}'' S3-4 emphasized the need for flexibility: ``\textit{there should be some flexibility in just being able to add our own situation right now, I think that is missing[...] whatever I will write, it will just be more generic. That will apply to everyone, which is not true.}'' These reflections highlighted participants' recognition that research lab contexts vary significantly depending on factors such as career stage, research area, and individual circumstances.
Providing personalized answers was expected to enhance \sys{}'s usefulness, pointing to the need for detailed, contextualized, and case-based content—beyond the lab’s formal policies in the repository.

\section{Discussion}

We reflect on our findings to understand implications for designing AI-mediated organizational memory systems. Our field study revealed several key tensions and opportunities that extend beyond \sys{}'s technical features to more fundamental questions about how knowledge management systems can support collaborative work in research organizations.

\subsection{Privacy-Awareness Tension in AI-mediated Organizational Memory}

Our findings revealed a fundamental tension between students' privacy preferences and managers' need for awareness in organizational memory systems. Because the majority of Q\&A took place through direct messages to \sys{} without being shared, directors had limited visibility into knowledge gaps in the documentation. Directors expressed willingness to update information that their students could not access through interactions with \sys{}. This challenge reflects a broader issue in collaborative work: maintaining awareness of individual and group activities is essential for successful collaboration~\cite{dourish1992awareness, tam2006framework}, yet our study shows that AI-mediated knowledge retrieval can suffer from the same limitation. 

Our analysis suggests that AI intermediaries such as \sys{} reconfigure organizational memory (OM) practices by redistributing visibility across members of an organization. By enabling workers to pose questions directly to an AI agent, these systems increase the visibility of existing organizational knowledge to questioners while reducing the workload of searching or manually locating information. However, this enhanced accessibility simultaneously diminishes members' organic awareness of knowledge gaps, which was previously supported through in-person interactions. This redistribution produces new asymmetries and hidden dynamics in a group, requiring members to negotiate trade-offs between individual privacy, exposure of information needs, and the collective benefits of shared organizational knowledge. Further, the interaction traces generated through AI-mediated OM—such as question-asking histories—can serve as a valuable resource for identifying missing knowledge or understanding evolving information needs within the organization. Yet leveraging such data also risks inadvertently creating forms of workplace surveillance, particularly when anonymity is contained in a smaller group. These dynamics reveal that AI-mediated OM systems do not merely support knowledge access; they actively reshape the visibility, accountability, and communication of knowledge sharing practice, offering a lens for understanding how algorithmic intermediaries influence sociotechnical dimensions of OM. 

This redistribution of visibility intersects with a long-standing tension in communication research between psychological safety and organizational transparency. The tension stems from what communication researchers have identified as the fundamental trade-off between psychological safety and organizational transparency. Students' behaviors—seeking private channels to avoid judgment and protecting others from perceived disruption—represent manifestations of psychological safety-seeking in hierarchical organizations. Research has consistently demonstrated that anonymous communication reduces social pressure and enables more honest self-expression, particularly when interacting with authority figures~\cite{suler2004online, clark-gordon2019anonymity}.  However, this creates what organizational behavior scholars have identified as a fundamental paradox: while privacy enables honest participation, complete transparency can also discourage participation by making members feel overly monitored~\cite{bernstein2012transparency}.

These findings suggest that future organizational memory systems must balance individual privacy with the members' benefit coming from reinforced organizational memory. Our participants suggested privacy-preserving awareness mechanisms, one of which is to periodically report questions at high levels, which would inform managers about question patterns without compromising individual privacy.  Given the students' preference for direction communication over \sys{}-mediated communication, this alternative could be more effective for lab directors to be aware of the gaps in organizational memory. This approach preserves student autonomy while providing managers with actionable insights about knowledge gaps and documentation needs.

\subsection{Expanding the Perceived Boundary of Documentable Knowledge}

Our study revealed a gap between the knowledge that lab members possessed and what they felt was appropriate to document. Many participants hesitated to share context-specific experiences that may be generalizable and beneficial to their peers.
This reluctance appears to stem from an implicit expectation that documentation should be universal, authoritative, and permanent—creating a high bar for contribution that excludes valuable experiential knowledge and partial solutions. Participants also recognized value in case-based information and partial policies that could provide personalized guidance when comprehensive documentation was unavailable.

This challenge is well-studied: from early work on organizational memory systems~\cite{conklin1997designing} to research on knowledge sharing barriers~\cite{ackerman2013sharing}, studies have consistently identified the difficulty of capturing informal knowledge due to cultural and technical barriers toward outcome-focused documentation. \newadd{Moreover, research has shown that knowledge needs in document work are inherently role-dependent and context-specific~\cite{jahanbakhsh2022understanding}, challenging assumptions of universal documentation.} Prior work on enterprise wikis found that members only occasionally contributed despite recognizing documentation value, often requiring explicit requests from managers~\cite{10.1145/1822258.1822283, kiniti2013wikis}. CHOIR addressed this barrier by embedding documentation opportunities within natural conversation flows. Participants noted that the knowledge extraction feature made contributing informal knowledge feel less burdensome, as it transformed ongoing discussions into documentation without requiring separate writing effort. This suggests that conversational AI systems can provide lightweight pathways for capturing experiential knowledge that traditional documentation systems have struggled to elicit.

Building on these findings, we propose a more expansive model of organizational memory that embraces contextual knowledge alongside formal policies. Future systems should support lightweight contribution mechanisms that allow members to share experiences, partial solutions, and case-based knowledge without the burden of creating universally applicable documentation. Furthermore, systems can provide guided pathways that help users navigate between formal documentation and informal human expertise when contributing knowledge. This concept aligns with early systems like Answer Garden~\cite{ackerman1990answer}. In addition, establishing common ground between lab directors and students, and fostering a welcoming environment even when the knowledge is canonical, may encourage students to contribute.

\subsection{AI Agents as Mediator in Organizations}

While recent research has demonstrated significant declines in human participation following AI introduction in knowledge communities~\cite{burtch2024consequences}, our findings demonstrate how AI-based chatbots can lead to markedly different outcomes for social dynamics within organizations. Rather than replacing human communication, CHOIR served as an intermediary that facilitated and preserved subsequent human interactions.
Evidence for this facilitative effect was particularly strong in L1, where public Q\&A usage corresponded with increased documentation activity, collaborative knowledge identification, and peer learning. Students explicitly used CHOIR to help identify missing documentation for their director, demonstrating how AI-mediated knowledge work can complement human collaboration. 

From a socio-technical perspective, CHOIR's mediating role can be understood through the lens of boundary objects~\cite{star1989institutional}. Organizational memory documents serve as boundary objects with interpretive flexibility, where directors view them as comprehensive repositories of lab knowledge that they want to convey to students, while students see them as practical guides for navigating their immediate questions and expectations. In our work, an AI agent mediates existing boundary objects to each group, and the AI agent itself is a new boundary object that each group can access and communicate with.
Extending recent work on AI systems as boundary objects in creative and participatory contexts~\cite{chung2023artinter, ayobi2021machine, porquet2025copying, guridi2025fake}, \sys{} shows how AI can serve this role in organizational memory, dynamically adapting the same knowledge base to serve different communities' needs.
This mediation extends organizational knowledge from context-dependent exchanges to persistent and reusable organizational memory that different members can retrieve and apply to their own situations.

These findings suggest an alternative to common framings of AI as either an assistant or a replacement for the human workforce. Our results point toward a model where AI serves as a facilitator that creates multiple channels for more effective human communication. This perspective aligns with prior work exploring conversational agents in collaborative settings~\cite{kim2020bot, shin2023introbot}. In our study, CHOIR reduced social barriers (through anonymity options), provided shared reference points for discussion (through document grounding), and surfaced implicit knowledge needs (through gap identification). The design implications extend beyond simple automation to consider how AI systems can actively foster human connection and collaboration. 

\subsection{Knowledge Access across Organizations} 

While our study focused on individual research labs, participants' needs frequently extended beyond their immediate organizational boundaries. Participants expressed frustration with CHOIR's limited access to external resources (e.g., a webpage on their department website) and desired connections to higher-level institutional documentation. This reflects a fundamental mismatch: users naturally seek knowledge at multiple levels and across the organization. Research on online communities demonstrates the value of such cross-organizational knowledge sharing, showing how similar governance structures across communities can provide valuable resources for new organizations lacking documentation~\cite{fiesler2018reddit, chandrasekharan2018internets, chandrasekharan2019crossmod}. This challenge would reflect the complexity found in other types of organizations, such as companies with many specialized teams performing different functions. Recent work on modular knowledge architectures offers promising technical approaches~\cite{zhao2025knoll}. While such approaches primarily address retrieval challenges, they point toward architectures that could support the complex permission structures, review workflows, and cross-boundary access patterns that hierarchical organizations require.

\subsection{Limitations}

Our study has several limitations that may impact the generalizability. First, our field deployment involved a small sample of participants across four university research labs, all from STEM fields with high technical literacy. This limits generalizability to organizations with different technical backgrounds, hierarchical structures, or cultural norms around knowledge sharing. Second, the one-month deployment period was sufficient for observing initial adoption patterns but insufficient for evaluating the longer-term effectiveness of organizational memory systems. Our study focused on discovering potential effects rather than proving sustained efficacy. 

Third, because the study was conducted during the summer, the findings primarily reflect the onboarding experiences of new remote students. Although this context provided insights into early-stage lab integration, usage patterns during regular academic semesters may differ. We plan to conduct a long-term follow-up with participating labs to examine if and how usage and perceptions evolve across different academic periods. Additionally, the study setup required preparing initial documentation for participating labs by creating new seed documents or converting existing materials into standardized formats. We acknowledge that these prepared documents may have affected ecological validity compared with naturally evolved documentation. 

Fourth, one participating lab (L4) was led by the corresponding author, which may introduce potential bias despite our efforts to minimize researcher influence. \newadd{L4 may also be particularly well aligned with the assumptions and practices that CHOIR supports, given the system's origin in addressing needs observed in this context.} While the corresponding author limited their role to managing the process, answering questions, and approving documents, and the first and second authors did not participate as study subjects, peer-pressure inherent in being in the same lab or the natural interest in a research project that is thematically similar to their own research could have influenced the L4 participants' engagement patterns or feedback. However, the key findings and discussion points are valid as they are not disproportionately driven by L4 data alone.

Lastly, our study lacks a controlled baseline and quantitative performance measures beyond descriptive statistics, which limits our ability to make causal claims about \sys{}'s effectiveness compared to alternatives such as wiki pages or existing lab practices (e.g., Google Docs/Slack Pinned Messages). As a result, our findings are descriptive rather than comparative, and further research is needed to determine how specific features of \sys{} contribute to its impact. \newadd{Future controlled experiments could compare an improved version of CHOIR against baseline conditions where shared documents and Slack are not connected through an AI mediator. Such studies could measure question response time, documentation update frequency and quality, knowledge accessibility, and participation in organizational memory maintenance.}

\section{Conclusion}

In this paper, we presented CHOIR, a chatbot-mediated organizational memory system that integrates document-grounded Q\&A, knowledge sharing, conversation-based knowledge extraction, and AI-assisted document updates within workplace communication platforms. Through a formative study with 15 participants, we identified key challenges in managing organizational memory in research labs, which guided our design of CHOIR. We then deployed CHOIR across four university research labs for one month, observing that it supported knowledge retrieval and documentation practices, with 42\% of questions answered from existing documentation and 1,994 words added to organizational memory. Our findings revealed a fundamental privacy-awareness tension in AI-mediated organizational memory systems, where students' privacy-seeking behaviors, driven by concerns about social evaluation, limited directors' awareness of knowledge gaps. We found that conversational agents can serve as facilitators rather than replacements for human communication, offering multiple channels for more effective human communication while identifying gaps in organizational documentation. Our work provides empirical observations and design implications for integrating AI-mediated organizational memory into workplace communication platforms, highlighting the importance of addressing psychological safety, power dynamics, and cultural norms in hierarchical organizations. These findings offer insights for designing knowledge management systems that balance individual privacy with organizational transparency and support capturing contextual knowledge that often remains \mbox{undocumented}.

\section{Acknowledgments}
We thank all professors and students who participated in the user studies, as well as the reviewers for their constructive feedback.
\bibliographystyle{ACM-Reference-Format}
\bibliography{references} 

\newpage
\onecolumn
\appendix

\newpage
\onecolumn

\section{System Prompts}\label{sec:appendix}

This appendix provides the complete system prompts used in CHOIR for reproducibility. Model configuration details are provided in Section 4.5.2.

\subsection{Question Answering Prompt}

The following prompt is used for document-grounded Q\&A (Feature 1). User names are replaced with pseudonyms before being sent to the API for privacy protection.

\begin{verbatim}
You are CHOIR, a helpful AI assistant for [Organization Name]. 
Think of yourself as a knowledgeable senior student or friendly professor who's always ready to help with questions.

I have access to the organization's documentation and knowledge base, 
so I can help you find information and provide guidance based on what we have documented.

Organization Information:
- Organization: [Organization Name]
- About: [Organization Description]
- Today's date: [Current Date]
- Workspace: [Workspace Name]

IMPORTANT: First, determine if you can answer the user's question based ONLY on the provided references. 
Do NOT use your general knowledge or make assumptions beyond what's explicitly stated in the references.

Then respond with a JSON object containing:
1. ``canAnswer'': true/false - whether the references contain sufficient information to answer the question
2. ``response'': your answer based on the references, or a friendly explanation that you couldn't find the information

Guidelines for answering:
- Be friendly, approachable, and helpful - like a senior colleague who genuinely wants to help
- Answer ONLY based on the provided references below - do NOT add general knowledge
- If multiple references contain conflicting information, prioritize the first reference in the list
- If you cannot answer based on the references, encourage the user to ask others or start a discussion to help improve 
  our documentation
- When you cannot fully answer a question, mention which related references you found and what information they contain
- Use a warm, academic tone - professional but not overly formal

==== REFERENCES ====
[Retrieved document chunks]

==== CONVERSATION HISTORY ====
[Previous messages in the thread]

==== CURRENT QUESTION ====
[User's question]

Analyze whether you can answer based on the documentation and provide your response as JSON.
\end{verbatim}

\subsection{Knowledge Extraction Prompt}

The following prompt is used to extract documentable knowledge from conversations (Feature 3). User names are replaced with pseudonyms before being sent to the API.

\begin{verbatim}
You are CHOIR, a documentation specialist. 
Extract organizational knowledge from the conversation that would be valuable for future reference.

Only extract information that establishes policies, procedures, or reusable knowledge for the organization. 
Do NOT extract personal preferences, individual decisions, or casual conversation.

Start with a descriptive markdown section title (using heading format with [Topic Name]), then write the information in natural paragraph format. 
Only include facts that are directly stated in the conversation - do not add explanations, interpretations, 
or implications.

Do not include personal conversation details like who said what. Always preserve any URLs mentioned.

Conversation:
[Conversation messages]

What information is shared in the conversation that should be documented?
\end{verbatim}

\subsection{Document Update Prompt}

The following prompt is used to integrate extracted knowledge into existing documentation (Feature 4).

\begin{verbatim}
You are a document editor. Improve existing content by integrating the provided knowledge.

Existing content type: [paragraph or list]

Rules:
- PRIORITIZE updating/replacing existing content when knowledge provides better, more accurate, 
  or more comprehensive information
- If knowledge contradicts existing content, prefer the knowledge (assume it's more current/accurate)
- If knowledge complements existing content without contradiction, add it in matching format
- If knowledge provides more specific details about existing points, merge them into improved versions
- Preserve all URLs from the knowledge
- Use single-level lists only (no nested bullets)
- No headings or section titles
- Return original only if knowledge adds no meaningful value

Approach: Update first, then add if needed. Create the most accurate and comprehensive version.

Wrap your response in <markdown> tags.

File: [File Name] - Section: [Section Name]

Existing content:
<markdown>
[Current section content]
</markdown>

Knowledge to integrate:
[Extracted knowledge]
\end{verbatim}

\end{document}